\documentclass[twocolumn,showpacs,preprintnumbers,amsmath,amssymb,superscriptaddress]{revtex4}
\usepackage{graphicx}
\usepackage{dcolumn}
\usepackage{bm}
\usepackage{longtable}
\usepackage[countmax]{subfloat}
\usepackage{epstopdf}

\providecommand{\be}{\begin{equation}}
  \providecommand{\ee}{\end{equation}}
\providecommand{\bea}{\begin{eqnarray}}
  \providecommand{\eea}{\end{eqnarray}}
\providecommand{\ba}{\begin{eqnarray}}
  \providecommand{\ea}{\end{eqnarray}}
\providecommand{\beas}{\begin{eqnarray*}}
  \providecommand{\eeas}{\end{eqnarray*}}

\providecommand{\beni}{\begin{equation*}}
  \providecommand{\eeni}{\end{equation*}}

\providecommand{\bw}{\begin{widetext}}
  \providecommand{\ew}{\end{widetext}}

\def\Var{\operatorname{Var}}

\def\s{\sigma}

\def\d{\delta}

\usepackage{epsfig}
\usepackage{amssymb}
\usepackage{amsmath}
\usepackage{amsthm}
\usepackage{latexsym}
\def\be{\begin{equation}}
\def\ee{\end{equation}}
\def\bc{\begin{center}}
\def\ec{\end{center}}

\def\be{\begin{equation}}
\def\ee{\end{equation}}
\def\bea{\begin{eqnarray}}
\def\eea{\end{eqnarray}}

\begin{document}

\preprint{APS/123-QED}

\title{Topological properties of hierarchical networks}

\author{Elena Agliari}
\affiliation{Dipartimento di Matematica, Sapienza Universit\`a di Roma, P.le A. Moro 2, 00185, Roma, Italy.}
\author{Adriano Barra}
\affiliation{Dipartimento di Fisica, Sapienza Universit\`a di Roma, P.le A. Moro 2, 00185, Roma, Italy.}
\author{Andrea Galluzzi}
\affiliation{Dipartimento di Matematica, Sapienza Universit\`a di Roma, P.le Aldo Moro 5, 00185, Roma, Italy.}
\author{Francesco Guerra}
\affiliation{Dipartimento di Fisica, Sapienza Universit\`a di Roma, P.le A. Moro 2, 00185, Roma, Italy.}
\affiliation{Istituto Nazionale di Fisica Nucleare, Sezione di Roma}
\author{Daniele Tantari}
\affiliation{ Centro di Ricerca Matematica Ennio De Giorgi, Scuola Normale Superiore di Pisa, Piazza dei Cavalieri, 3, Pisa.}
\author{Flavia Tavani}
\affiliation{Dipartimento di Scienze di Base e Applicate per l'Ingegneria, Sapienza Universit\`a di Roma, Via A. Scarpa 16, 00185, Roma, Italy.}
\date{\today}

\begin{abstract}
Hierarchical networks are attracting a renewal interest for modelling the organization of a number of biological systems and for tackling the complexity of statistical mechanical models beyond mean-field limitations. Here we consider the Dyson hierarchical construction for ferromagnets, neural networks and spin-glasses, recently analyzed from a statistical-mechanics perspective, and we focus on the topological properties of the underlying structures. In particular, we find that such structures are weighted graphs that exhibit high degree of clustering and of modularity, with small spectral gap; the robustness of such features with respect to the presence of thermal noise is also studied. These outcomes are then discussed and related to the statistical mechanics scenario in full consistency.
Lastly, we look at these weighted graphs as Markov chains and we show that in the limit of infinite size, the emergence of ergodicity breakdown for the stochastic process mirrors the emergence of meta-stabilities in the corresponding statistical mechanical analysis.
\end{abstract}
 \maketitle


\section{Introduction}

When dealing with statistical-mechanics models (e.g. spin systems), overcoming the mean-field approximation is extremely challenging. Basically, the mean-field approximation lies in the assumption that each spin $S_i$ ($i=1,...,N$) in an embedding space does interact with all the other $N-1$ spins with the same strength, notwithstanding their mutual distance, as if spins occupied the $N$ vertices of
an hyper-tetrahedron.
As a notion of distance is introduced and couplings among spins are accordingly rescaled, the exact solution is, in most cases, out of reach. 

In the 60's, a hierarchical model for ferromagnetic systems was introduced to
describe non-mean-field spin systems \cite{dyson}, and it is known as the Hierarchical Ferromagnet.
More recently, also the Sherrington-Kirkpatrick model for spin-glasses \cite{castellana-prl,castellana-pre,castellana-jsp} and the Hopfield model for neural networks \cite{NOI-PRL, NOI-JPA, NOI-NN} defined on such a hierarchical topology have been investigated.

The hierarchical network exploited in all these cases is endowed with a metric and it is explicitly not-mean-field since the coupling between two nodes at a distance $d$ scales as $\sim 4^{-\sigma d}$, where $\sigma$ is a proper tuneable parameter. As a result, the spins can be thought of as placed on the vertices of a fully-connected weighted graph, where the coupling pattern mirrors the mutual distance among spins.
This graph exhibits peculiar features (e.g., high degree of modularity), which play a crucial role in the statistical-mechanics treatability as well as in the emergent behavior of the above mentioned models.
Also, the knowledge of the specific architecture considered allows to figure out the class of real-world systems where theoretical results can properly be applied.
However, only marginal attention has been devoted to such topological properties in the past and in this work we just aim to deepen these aspects.

In the following we first provide a streamlined and general introduction to the statistical-mechanics models considered (i.e., the hierarchical ferromagnet, the hierarchical neural network and the hierarchical spin glass), then we move to the analysis of the underlying network by studying the degree of clustering, the modularity, the ergodicity and the spectral properties. Finally, a section with outlooks and conclusions closes the paper.

\begin{figure*}\label{fig:albero}
	\includegraphics[width=9cm]{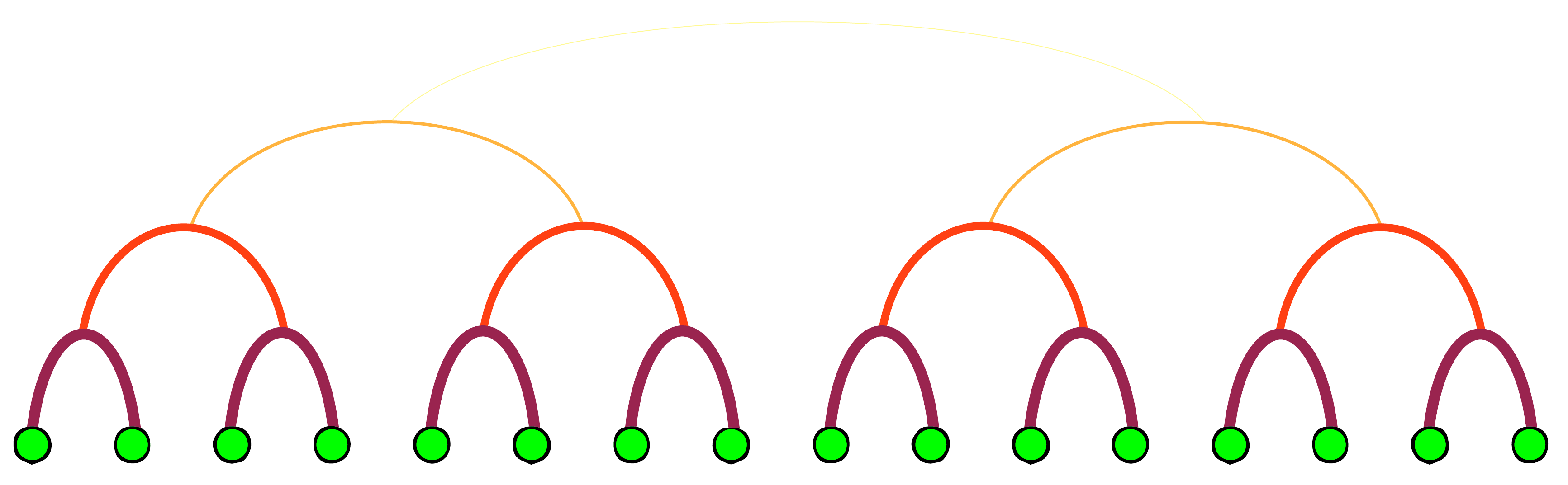}
	\caption{(Color online) Schematic representation of the hierarchical topology that underlies
		the system under study: spots represent nodes where spins/neurons live,
		while different colors and thickness for the links mimic different intensities in
		their mutual interactions: the brighter and thinner the link, the smaller the
		related coupling.}
\end{figure*}

%

\section{Definition of models and related Hamiltonians}
The three statistical-mechanics models which we adapt to live on a hierarchical network are the Curie-Weiss model, the Hopfield model and the Sherrington-Kirkpatrick model, which are the prototypes for ferromagnetism, associative neural networks and spin-glasses, respectively.

Before providing the Hamiltonians of these models when defined in a hierarchical structure, we outline how they can be built up recursively.
One starts from a set of two spins properly coupled (the kind of coupling depending on the particular model considered). Then, one takes two of such dimers and makes two operations: update the existing links and introduce new links to couple spins belonging to different dimers. This constitutes the system at the first iteration. At the next iteration, one takes two replicas of  such a system and, again, updates the existing links and introduces new connections among spins from different replica and so on up to the $k$-th iteration. In this way one can immediately see that a notion of distance emerges straightforwardly as two spins can be considered at a distance $d$ if they are first connected at the $d$-th iteration (see Figs.~$1$ and $2$).

More formally, the hierarchical ferromagnet (HFM) with $K$ levels of iterations is described by the Hamiltonian $H_{K}^{\textrm{HFM}}$, defined recursively as
\begin{eqnarray} \label{dyson}
\nonumber
H_{K}^{\textrm{HFM}}(\{ S \} |\sigma) &=& H_{K-1}^{\textrm{HFM}}(\{S_1 \}|\sigma) + H_{K-1}^{\textrm{HFM}}(\{ S_2 \}|\sigma)\\
&-& \frac{1}{2^{2\sigma K}}\sum_{i<j}^{2^{K}} S_i S_j,
\end{eqnarray}
where $\{ S \}$ is the set of $N=2^K$ spins making up the system, each labeled as $i=1,...,N$, while $\{ S_1\}$ and $\{ S_2\}$ are the sets of spins related to the two smaller copies of sizes $2^{K-1}$ that are merged up. Spins are binary and can take values $+1$ or $-1$. The parameter $\sigma$ is bounded as $\sigma \in (1/2, \ 1]$: for $\sigma >1$ the interaction energy goes to zero in the thermodynamic limit, while for $\sigma< 1/2$ the interaction energy diverges in the same limit. Also, notice that the coupling among spins is positive due to the ferromagnetic nature of the model which makes neighboring spins to ``imitate'' each other.

Next, the Hopfield model requires for its definition the set of $N$ quenched vectors $\{ \xi^{i} \}$, $i=1,...,N$, of length $P$ and whose entries are drawn from the distribution
\be \label{eq:attribute}
P(\xi_i^{\mu})=\frac12 \delta(\xi_i^{\mu}-1) + \frac12 \delta(\xi_i^{\mu}+1),
\ee
being $\mu =1,...,P$. By applying the Mattis gauge $S_i\rightarrow -S_i\xi_i^{\mu}$, and summing over the $P$ patterns, the Hamiltonian $H_{K}^{\textrm{HNN}}$ for the hierarchical neural network (HNN), at the $K$-th level of iteration, reads as
\begin{eqnarray}\label{HHM}
\nonumber
H_{K}^{\textrm{HNN}}(\{ S \}|\xi,\sigma) &=& H_{K-1}^{\textrm{HNN}}(\{ S_1 \} |\xi,\sigma) + H_{K-1}^{\textrm{HNN}}(\{ S_2 \}|\xi,\sigma) \\
&-& \frac12 \frac{1}{2^{2\sigma K}}\sum_{\mu=1}^P\sum_{i,j=1}^{2^{K}}\xi_i^{\mu}\xi_j^{\mu} S_i S_j,
\end{eqnarray}
with $H_0^{\textrm{HNN}}\equiv 0$ and $\sigma$ still within the previous bounds, i.e. $\sigma \in (\frac{1}{2}, 1]$.

Finally, the hierarchical spin-glass (HSG) requires for its definition the set of $N(N-1)/2$ quenched variables $\chi_{ij}$ drawn from a standard centered Gaussian distribution $\mathcal{N}[0,1]$ such that the related Hamiltonian $H_{K}^{\textrm{HSG}}$, at the $K$-th level of iteration, reads as
\begin{eqnarray}\label{SK}
\nonumber
H_{K}^{\textrm{HSG}}(\{ S \}|\chi,\sigma)&=&  H_{K-1}^{\textrm{HSG}}(\{ S_1 \}|\chi,\sigma) + H_{K-1}^{\textrm{HSG}}(\{ S_2 \}|\chi,\sigma)\\
& -& \frac{1}{2^{2\sigma K}}\sum_{i<j}^{2^{K}}\chi_{ij} S_i S_j,
\end{eqnarray}

All these models (i.e., HFM, HNN, HSG) can be thought of as spin systems embedded on a weighted graph $\mathcal{G}=(V,E)$, where $V$ is the set of nodes labeled as $i=1,\cdots,2^K$ and $E$ is the set of links whose cardinality is $|E|=2^{K-1}(2^K-1)$. Each spin $S_i$ occupies the vertex $i \in V$ and each link $(i,j) \in E$ is associated to a weight $J_{ij}$ capturing the effective coupling among spins. Then, in general, the Hamiltonians in (\ref{dyson}), (\ref{HHM}), and (\ref{SK}) can all be written in the compact form
\be
H_{K}^{\textrm{(model)}}(\{S\}|\sigma) = \sum_{ij} J_{ij}^{(\textrm{model})} S_i S_j.
\ee



\begin{figure*}\label{fig:grafo}
\includegraphics[width=9cm]{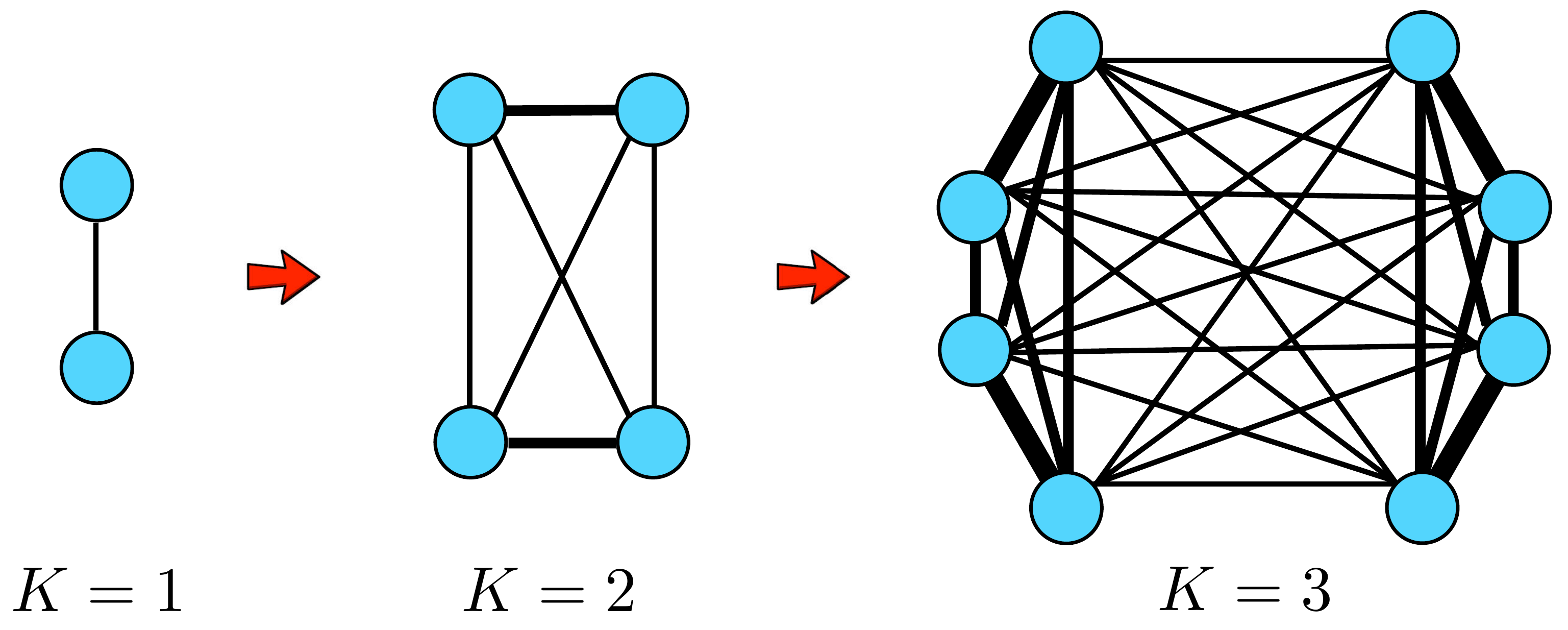}
\caption{(Color online) Iterative construction of the hierarchical structure up to generation $K=3$, corresponding to $N=2^3$ vertices. Links display different thickness according to their weight.}
\end{figure*}

\section{Graph generation in the hierarchical ferromagnet} \label{sec:GD}
In this section we focus on the generation of the weighted graph $\mathcal{G}$ underlying the Hamiltonian $H_{K}^{\textrm{HFM}}(\{ S \}|J,\sigma)$ and in the next subsections we will analyze its properties.

The iterative construction outlined in the previous section can be adopted to build up $\mathcal{G}$ (see Fig.~$2$): we start from a couple of nodes $i$ and $j$, connected by a link carrying a weight $J_{ij} = J_1(1) = 4^{-\sigma}$, and we refer to this graph as $\mathcal{G}_1$. Then, we take two replicas of $\mathcal{G}_1$ and we connect nodes belonging to different replicas with links carrying a weight $J_2(2) = 4^{-2\sigma}$, while existing links are updated as $J_2(1) = J_2(1) + J_2(2) = 4^{-\sigma} + 4^{-2\sigma}$. The graph $\mathcal{G}_2$ therefore counts $2^2$ nodes.  We proceed iteratively in such a way that at the $K$-th iteration new links connecting nodes belonging to different replicas  are associated to a weight $J_K(K) = 4^{-K\sigma}$, while existing links in each replica $\mathcal{G}_{K-1}$ are all increased by the same value $J_K(K)$, in such a way that
\be \label{eq:Jdef}
J_K(d) =\sum_{l=d}^{K} J_K(l) = \sum_{l=d}^{K} 4^{-l \sigma} = \frac{4^{\sigma(1-d)} -4^{-K \sigma}}{4^{\sigma} -1}.
\ee
The resulting graph $\mathcal{G}_K$ (simply referred to as $\mathcal{G}$ to lighten the notation), is undirected and fully connected. Its nodes make up a set $V$ of size $N=2^K$ and are labeled as $i=1,...,N$.
Also,  the set of links $E$ contains all possible $\binom{N}{2}$ connections as the graph is fully connected, and each link $(i,j) \in E$ is associated to a weight $J_{ij}$ which can be defined in terms of the distance between nodes $i$ and $j$, once a proper metric has been introduced.

In fact, the procedure described above provides a notion of distance $d$ which we recall here: two nodes are said to be at distance $d$ if they are first connected at the $d$-th iteration. For completeness, we also fix $\mathcal{G}_0$ as the graph consisting in a single node.

As a result, this metric is intrinsically ultrametric as, for any pair $i,j \in V$, we have

-$d_{ij} \geq 0;$  \\
$\mbox{ }\mbox{ }\mbox{ }$-$d_{ij} = 0 \text{ iff } i=j;\mbox{  }$\\
$\mbox{ }\mbox{ }\mbox{ }$-$d_{ij} = d_{ji}, \text{that is, the metric is symmetric};$\\
$\mbox{ }\mbox{ }\mbox{ }$-$d_{ij} \leq \textrm{max} (d_{iz}, d_{zj})$ (this is the so-called ultrametric inequality).


Beyond the definition of distance $d_{ij}$ based on recursivity, we can straightforwardly adopt the $p$-adic metric \footnote{
We define the $p$-adic metric in terms of the p-adic norm exactly the way that we defined Euclidean distance in terms of the absolute value norm. In the P-adic integers, the norm of a number is based around the largest power of the base that's a factor of that number: for an integer $x$, if $p^n$ is the largest power of $p$ that is a factor of $x$, then the the p-adic norm of $x$ (written $||x||_p$) is $p^{-n}$. So the more times you multiply a number by the $p$-adic base, the smaller the $p$-adic norm of that number is.
The way we apply that to the rationales is to extend the definition of p-factoring: if  is our $p$-adic base, then we can define the $p$-adic norm of a rational number as: $i.$ $||0||_p =0$, $ii.$ For other rational numbers $x$: $||x||_p=p^{-\textrm{ord}_p(x)}$:  where, if $x$ is a natural number, then $ord_p(x)$ is the exponent of the largest power of $p$ that divides $x$. Therefore, two $p$-adic numbers $x$ and $y$ are close together if  $x-y$ is divisible by a large power of $p$.
Closest nodes are at distance $2^{-K+1}$, farthest nodes are at a distance $1$.
}, and measure the $p$-adic distance $\rho_{ij}$ between nodes $i$ and $j$, as (here $p$ is set equal to $2$)
\be
\rho_{ij} = || i- j||_2 = 2^{-\textrm{ord}_2(i-j)},
\ee
being $\textrm{ord}_2(i-j)$ the exponent of the largest power of $2$ that divides $(i-j)$ \footnote{This alternative definition of distance can also be useful for the numerical implementation of the hierarchical graph under study.}. Notice that $\rho$ is connected with $d$ by $d_{ij} = K - \textrm{ord}_2(i-j)$.
As a result, $\rho_{ij} \in \{ 2^{-K+1}, 2^{-K+2}, ..., 2, 1 \}$. Then, using the $2$-adic metric, one can see that the coupling strength turns out to decay algebraically with the ($2$-adic) distance, as typical for long-range interactions, that is 
\be\label{eq:ABrho}
J_{ij} = \frac{A}{\rho_{ij}^{2\sigma}} + B.
\ee
In fact, by posing $A= \frac{4^{- \sigma (K-1)}}{4^{\sigma}-1} $ and $B = - 4^{-\sigma} A $, we recover the definition in (\ref{eq:Jdef}).
Moreover, 
we can rearrange Eq.~\ref{eq:Jdef} and, in the limit of large size, we get
\be \label{eq:J}
J_{ij} = \frac{2^{-2 \sigma K}}{4^{\sigma}-1} \left[ \left( \frac{2}{\rho_{ij}} \right)^{2 \sigma} -1 \right]  \sim \frac{1}{(N \rho_{ij})^{2 \sigma} }.
\ee
The two extrema for $\sigma$, i.e. $\sigma=1/2$ and $\sigma=1$, therefore correspond to a coupling strength scaling linearly and quadratically, respectively, with the ($2$-adic) distance between nodes.

As anticipated, the hierarchical ferromagnet in (\ref{dyson}), is obtained by pasting on each vertex $i$ a spin $S_i$ and letting spins interact with a coupling $J_{ij}$.

The formalization just described can be properly extended to allow for a degree of stochasticity, e.g. the set of labels can be extracted from a suitable distribution, and/or $p$ can be varied hence generating structures based on $p$-plets rather than on couples, that is, ultimately hierarchical $p$-spin models \cite{castellana-jsp2} or their $p\to \infty$ limit known as hierarchical random energy model \cite{castellana-prl}. Here we focus on the deterministic case depicted in Fig.~$2$, which holds for pairwise interactions only.

\begin{figure}\label{fig:distr}
\includegraphics[width=7cm]{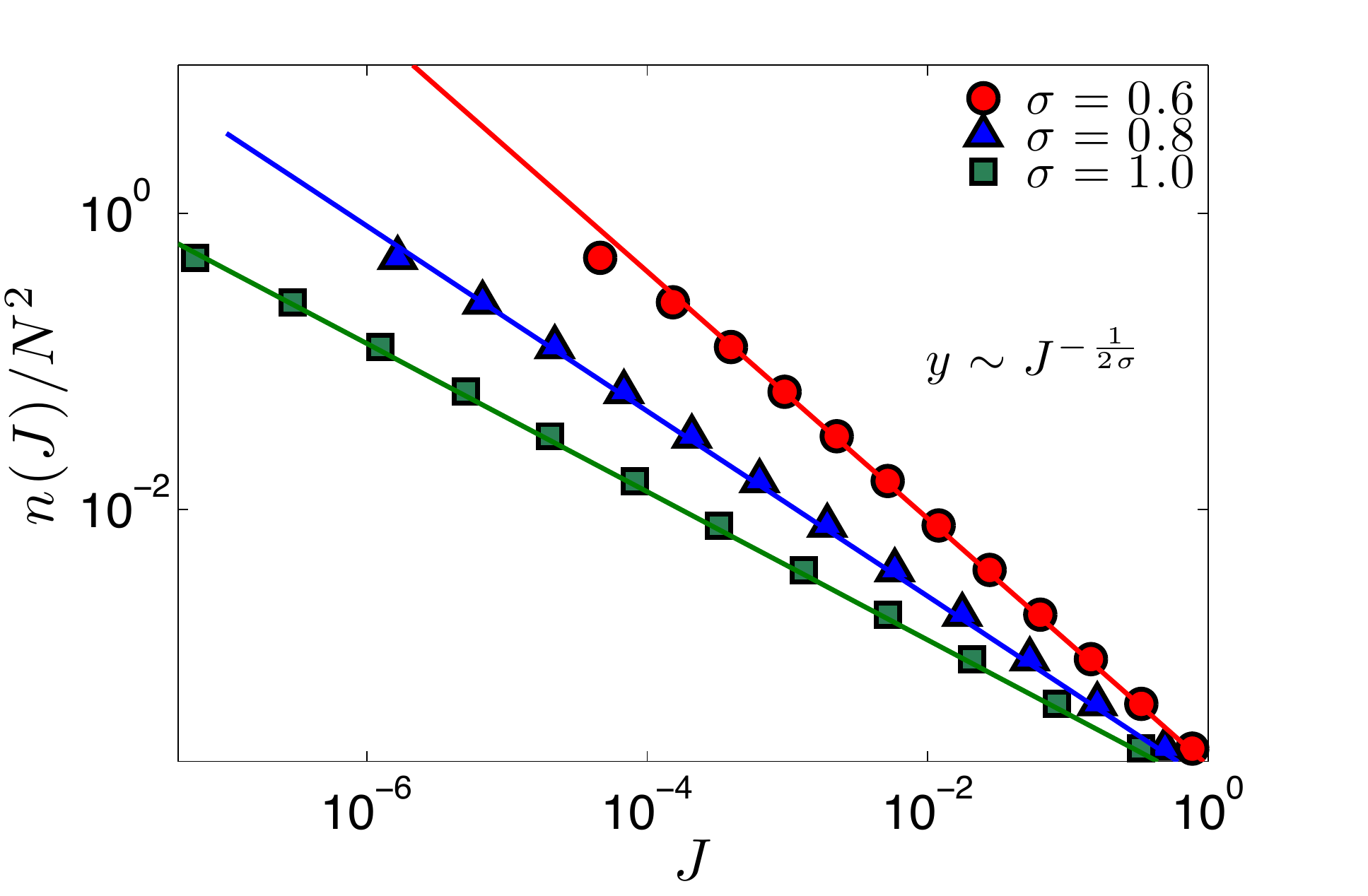}
\caption{(Color online) Distribution $n_k(J)$ for $K=12$ and $\sigma=0.6, \sigma=0.8$, and $\sigma=1.0$, as reported by the legend. Straight lines correspond to $y=J^{-1/(2\sigma)}/(2N)$, see Eq.~\ref{eq:nJ}. }
\end{figure}

 We proceed the investigation by deriving the number $n_{K}(J)$ of links carrying weight $J$, which provides a picture of how weights are distributed in between the two extrema
\begin{eqnarray} \label{eq:co}
J_{\textrm{max}} &=& J(1) = \frac{1 - 4^{-K \sigma}}{4^{\sigma} -1}, \\
\label{eq:lb}
J_{\textrm{min}} &=& J(K) = 4^{-K \sigma}.
\end{eqnarray}
To this aim it is convenient to count the number $n(d)$ of couples $(i,j)$ such that $d_{ij}=d$ and which are therefore connected by a link with weight $J_{ij}=J(d_{ij})=J(d)$. In fact, we have
\be \label{eq:nJ1}
n_K(d)= 2^{d-1} \frac{N}{2} = 2^{K+ d -2},
\ee
and, of course, $\sum_{d=1}^K n(d)/\binom{N}{2}=2^{K-1}(2^K-1)/\binom{N}{2}=1$.

Moreover, by inverting the formula in Eq.~\ref{eq:Jdef}, i.e. $d = -1/(2 \sigma) \log_2 [(2N)^{-2\sigma} +J(1-2)^{-2 \sigma}]$, we can express $n(d)$ in terms of $J$, namely
\be \label{eq:nJ}
n_K(J)= \frac{N^2}{2} \left[ 1+J N^{2 \sigma} (2^{2 \sigma} -1) \right]^{-\frac{1}{2\sigma}} \approx \frac{N}{2} J^{-\frac{1}{2\sigma}},\\
\ee
where the last approximation holds for $N$ large and highlights that the distribution is power-law (although with cut-offs given by Eqs.~\ref{eq:co} and \ref{eq:lb}). Otherwise stated, this model can be looked at as a ``scale-free Curie-Weiss''.
%
The distribution $n_K(J)$ is depicted in Fig.~$3$, where different choices of $\sigma$ are as well compared, while in Fig.~$4$ the overall pattern of weights $\mathbf{J}$ is shown.

Another observable, closely related to the coupling matrix $\mathbf{J}$  is the weighted degree $w$ \cite{jtb2}. Differently from the (bare) degree $z$, which simply counts the number of links stemming from a node, the weighted degree also accounts for the weights associated to stemming links. More precisely, the weighted degree $w_i$ of node $i$ is defined as \be w_i = \sum_{j=1,j \neq i}^N J_{ij};\ee of course, since there is perfect homogeneity within this system $w_i \equiv w, \forall i$. From a statistical mechanics perspective, $w_i$ (respectively $-w_i$) represents the field acting on the $i$-th spins when all the remaining spins are pointing upwards (respectively downwards).
Recalling Eqs.~\ref{eq:Jdef} and  \ref{eq:nJ1}, we get
\begin{eqnarray} \label{eq:W}
&& w_K(\sigma) = \sum_{d=1}^{K} 2^{d-1} J(d)=\nonumber\\
&=& \frac{1}{2(4^{\sigma}-1)} \left[ 4^{\sigma} \sum_{d=1}^K 2^{d(1-2\sigma)} -4^{-K \sigma} \sum_{d=1}^K 2^d  \right]\label{eq:wpartial}\\
&=& \frac{1}{2(1-4^{-\sigma})} \left[ \frac{1-(2^{2\sigma-1})^{-K}}{2^{2\sigma-1}-1} -  \frac{(2^K-1)}{2^{2\sigma(K+1)} }\right]  \nonumber\\
&=& \frac{4^{\sigma}}{2(4^{\sigma} -1)} \left[ \frac{(2N)^{2 \sigma} -N(3 \times 2^{2 \sigma-1} -1) +2^{2\sigma} -1  }{(2N)^{2\sigma}(2^{2 \sigma-1}-1)}  \right],\nonumber\\ 
\end{eqnarray}
where in the first line $2^{d-1}$ is the number of neighbors at distance $d$. 
When $\sigma >1/2$, in the thermodynamic limit, we get
\be
w_K(\sigma>1/2)  \xrightarrow {K\gg1} \frac{4^{\sigma}}{(4^{\sigma}-1)(4^{\sigma}-2) }. 
\ee
It is worth stressing that, in the thermodynamic limit, the weighted degree $w_K(\sigma>1/2)$ remains finite, although the bare degree of any node goes to infinity.
On the other hand, when $\sigma=1/2$, using ($\ref{eq:wpartial}$), the first term in square brackets converges to $K-1$, while the second term converges to $1/2$, whence we have
\be
w_K(\sigma=1/2) =  \left[ K-1 - \frac{N-2}{2N} \right] \sim K,
\ee
that is, in the thermodynamic limit, the weighted degree has a logarithmical divergence with $N$ (we recall that $N=2^K$); coherently, the case $\sigma=1/2$ is excluded from the statistical-mechanics investigations \cite{NOI-NN,NOI-JPA}.
%

\begin{figure}\label{fig:carpet}
\includegraphics[width=6.5cm]{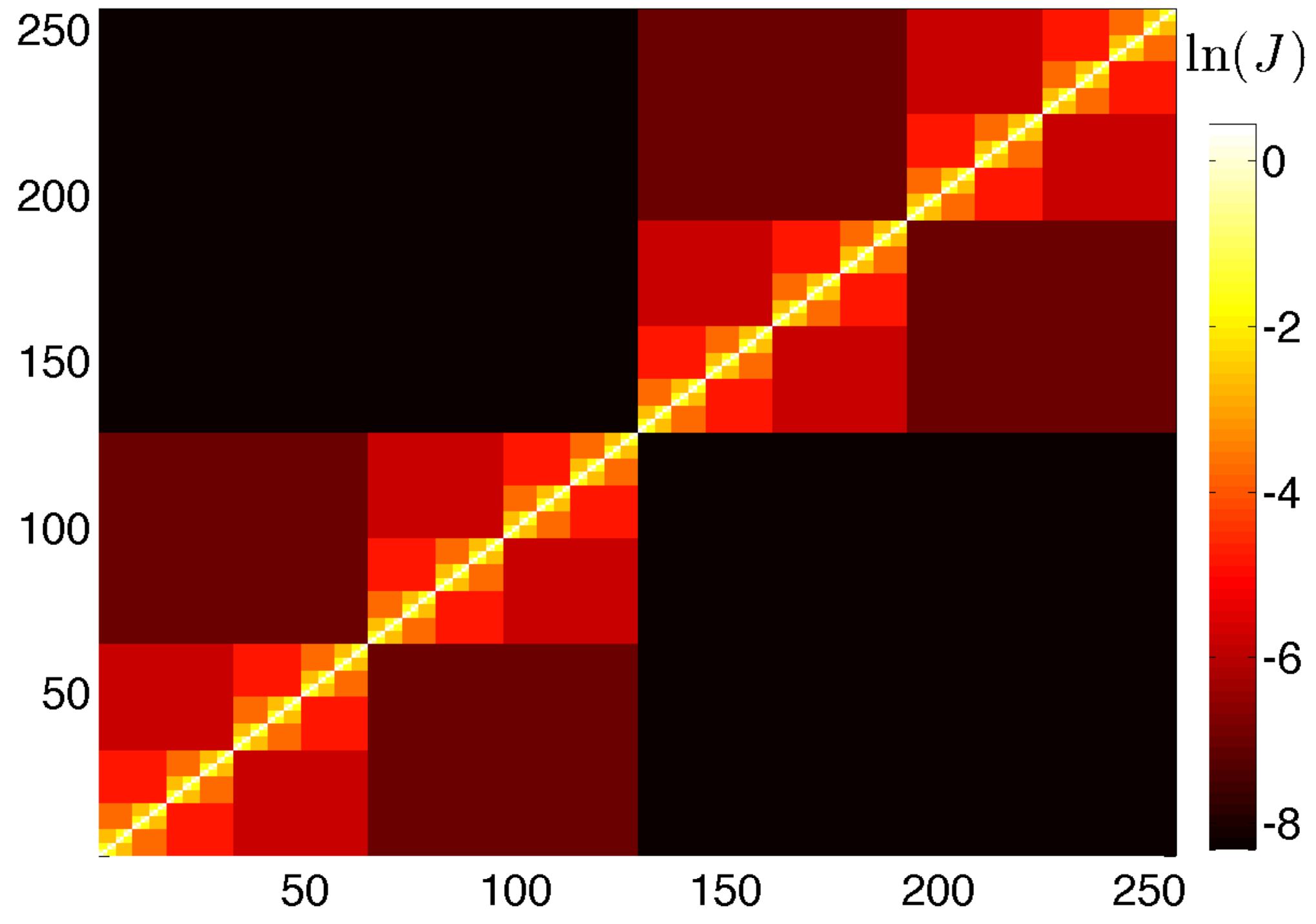}
\caption{(Color online) The pattern of weights $J_{ij}$ for $K=8$ is represented in a logarithmic scale. Notice that these  patterns mirror the ultrametric structure of the graph. 
}
\end{figure}

\begin{figure*}
\includegraphics[width=15cm]{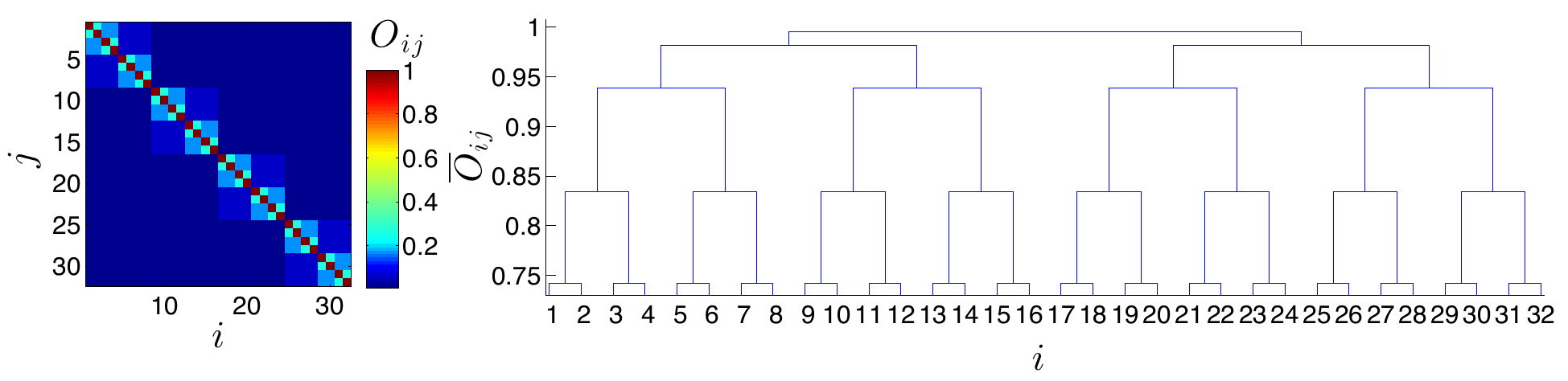}
	\caption{(Color online) Modularity of the graph $\mathcal{G}$ with $K=5$, $\sigma=0.9$, represented by means of the overlap matrix $\textbf{O}$ (left panel). Its regular structure  mirrors the deterministic modularity of the network. The dissimilarity of the graph is depicted through a dendrogram (right panel), where nodes at distance $d_{ij}=1$ (see for instance node $i=1$ and $j=2$, at the lower level of the dendrogram) have higher values of overlap (hence have lower values of dissimilarity) with respect to those at distance $d_{ij}=2$ (second level from the bottom of the dendrogram), up to the maximum distance $d_{ij}=5$ (first level on the top), underlying again the ultrametric structure of the network.}
	\label{fig:mod}
\end{figure*}

The last part of this section is devoted to the study of the network modularity and clustering.
Of course, when looking at the bare topology of the hierarchical network we have a fully-connected graph with no community structure and a trivial, unitary clustering coefficient. However, when weights on links are also taken into account one can highlight the emergence of a high degree of modularity and of clustering by properly extending the formula meant for unweighted networks. In particular, modularity can be quantified in terms of the generalized topological overlap matrix $\textbf{O}$ \cite{GTOM}, whose entry $O_{ij}$ measures the degree of similarity displayed by the couple of nodes $(i,j)$ in terms of the number of shared neighbours, namely
\be
O_{ij} = \frac{|N(i)\cap N(j)|+A_{ij}}{ \min \{ |N(i)|, |N(j)| \} -A_{ij} +1}, \label{eq:GTOM}
\ee
where $N(i)$ and $N(j)$  are the sets of nearest neighbors of $i$ and $j$, respectively, $|N(i)\cap N(j)|$ represents the number of common neighbours that nodes $i$ and $j$ share, and $\mathbf{A}$ is the adjacency matrix. 
Now, the presence of weights can be accounted for by modifying  Eq.~\ref{eq:GTOM} as proposed in \cite {Zhang-GMB2005}
\be  \label{eq:GTOM2}
O_{ij}= \frac{1}{J_{\textrm{max}}}  \frac{ \sum_{k=1}^{N} J_{ik} J_{kj} + J_{ij} J_{\textrm{max}}}{ \min \{ w_i, w_j \} - J_{ij} +J_{\textrm{max}}}.
\ee
Of course, Eq.~\ref{eq:GTOM2} recovers Eq.~\ref{eq:GTOM} as long as we replace the adjacency matrix with the normalized coupling matrix ($0 \leq J_{ij}/ J_{\textrm{max}} \leq 1$). The generalized topological overlap matrix for the graph under study is shown in 
the left panel of Fig.~\ref{fig:mod}, where one can see that $\textbf{O}$ mirrors the ultrametric structure of the graph.


Moreover, we can compute the degree of dissimilarity as
\be
\overline{O}_{ij}=1- O_{ij},\label{eq:disGTOM}
\ee
which is shown through a dendrogram plot in the right panel of Fig.~\ref{fig:mod}. Again, the ultrametric structure of the graph emerges markedly.
Further details on modularity can be found in Appendix A.

%
As for the clustering coefficient, several definitions of weighted clustering coefficient $cw$ have appeared in the literature, as summarised and compared in \cite{Saramaki-PRE2005}. Since there is not any ultimate formulation, we consider two definitions, introduced in \cite{Barrat-PNAS2004} and in \cite{Onnela-PRE2003}, respectively, which can be seen as limiting cases. According to the formula given in \cite{Barrat-PNAS2004}, we get
\be \label{eq:c1}
 cw_i^{(1)} = \frac{1}{w_i (z_i -1)} \sum_{j,h \in \mathcal{T}_i} \frac{J_{ij} + J_{ih}}{2},
\ee
where $\mathcal{T}_i$ is the number of triangles including node $i$, $w_i$ is the weighted degree of node $i$, $z_i$ is the number of nearest-neighbors of $i$ (i.e., its bare degree), and the normalization factor $w_i (z_i -1)$ ensures
 that $0 \leq cw_i^{(1)} \leq 1$. This definition of weighted clustering coefficient considers
only weights of edges adjacent to node $i$, but not the weights of edges between the neighbors of the
node $i$ (i.e. $J_{jh}$ in the previous formula).

Of course, the formula (\ref{eq:c1}) recovers the standard definition of clustering coefficient $c_i$ for unweighted graphs, namely $cw_i^{(1)} \to c_i$ as long as $J_{ij} \to 1$. Also, for the hierarchical graph considered here, due to homogeneity, $cw_i^{(1)}$ is node independent and can be simplified as
\begin{eqnarray}
\label{eq:c12}
\nonumber
 cw^{(1)} &=& \frac{1}{w (N-2)} \Big[\sum_{d=1}^K \sum_{\substack{d^{\prime}=1\\d^{\prime}\neq d}}^K  \frac{J(d)+J(d^{\prime})}{2}  2^{d-1} 2^{d^{\prime}-1}\\
 &+& 2 \sum_{d=1}^K J(d) \binom{2^{d-1}}{2} \Big] = 1.
\end{eqnarray}
The result in eq. (\ref{eq:c12}) derives from the fact that the hierarchical graph is fully connected, thus, as only weights of adjacent links are counted, the summation simply returns the weighted degree times the number of triangles including a given edge.

According to the definition given in \cite{Onnela-PRE2003}, we have
\be \label{eq:c2}
 cw_i^{(2)} = \frac{1}{(N-1)(N-2)} \frac{1}{J_{\max}} \sum_{j,h} \left( J_{ij} J_{ih} J_{jh} \right)^{1/3},
\ee
which is again normalized, i.e., $0 \leq cw_i^{(2)} \leq 1$, but, different from Eq.~\ref{eq:c1}, takes into account the weights of all edges making up a triangle and is invariant to weight permutation for one triangle.
As already noticed, due to the homogeneity of the graph under study, the clustering is node independent and hereafter we shall simply refer to  $cw^{(2)}$, dropping the index $i$.

With some algebra, we can rewrite the previous formula in terms of distances between nodes as
\beas
cw^{(2)} &=&  \frac{2}{(N-1)(N-2)}\frac{1}{J_K(1)}\sum_{d=1}^{K-1}3\times 2^{d-3}[J_K(d)]^{\frac{1}{3}}\times\nonumber\\
&\times & \sum_{d'=d+1}^{K}2^{d'}J_K(d')^{\frac{2}{3}},
\eeas
where $3 \times  2^{d+d'-3}$ is the number of triangles having two nodes at distance $d'$ from a fixed node, and being themselves at distance  $d$ each other.
Sobstituting $J_K(d)$ and $J_K(d')$ with their exact values given by $(\ref{eq:Jdef})$ and then assuming $K \gg 1$
\beas
(J(d)J(d')^2)^{\frac{1}{3}}&=& \Big[\frac{4^{\sigma(1-d)}-4^{-K\sigma}}{4^{\sigma}-1}\Big]^{\frac{1}{3}}\Big[\frac{4^{\sigma(1-d')}-4^{-K\sigma}}{4^{\sigma}-1}\Big]^{\frac{2}{3}}\approx\nonumber\\
&\approx & \frac{1}{4^{\sigma}-1}\Big[4^{\sigma(1-d)}4^{2\sigma(1-d')}\Big]^\frac{1}{3},
\eeas
we arrive to the following approximation of the clustering coefficient $cw^{(2)}$
\bea
&&cw^{(2)}\approx  \widetilde{cw}^{(2)} = \frac{3}{4(N-1)(N-2)}\frac{4^{\sigma}}{(4^{\sigma}-1)J_K(1)}\times\nonumber\\&\times &\sum_{d=1}^{K-1}\sum_{d'=d+1}^{K}2^{d(1-\frac{2}{3}\sigma)}2^{d'(1-\frac{4}{3}\sigma)}=\\
\label{eq:Onn}
&=& \frac{3(2N)^{2\sigma}[2^{\frac{2\sigma}{3}+1}-4+2^{\frac{2\sigma}{3}}N^{2(1-\sigma)}(2^{\frac{4\sigma}{3}}-2)]}{(2^{\frac{2\sigma}{3}}-2)(2^{\frac{4\sigma}{3}}-2)(2-3N+N^2)(2^{2\sigma}-4)(N^{2\sigma}-1)}+\nonumber\\
\nonumber
&-&\frac{3 N^{\frac{2}{3}\sigma+1} 2^{2\sigma}(2^{2\sigma}-4)}{(2^{\frac{2\sigma}{3}}-2)(2^{\frac{4\sigma}{3}}-2)(2-3N+N^2)(2^{2\sigma}-4)(N^{2\sigma}-1)}.
\eea
This approximation provides the leading behavior for $cw^{(2)}$ in the limit of large size. It is worth noticing that, differently from the previous definition (\ref{eq:c1}),  here $cw^{(2)}$ is always close to zero, due to presence in the graph of a high number of triangles constituted by distant nodes.

\begin{figure}[h!]
\includegraphics[width=8cm]{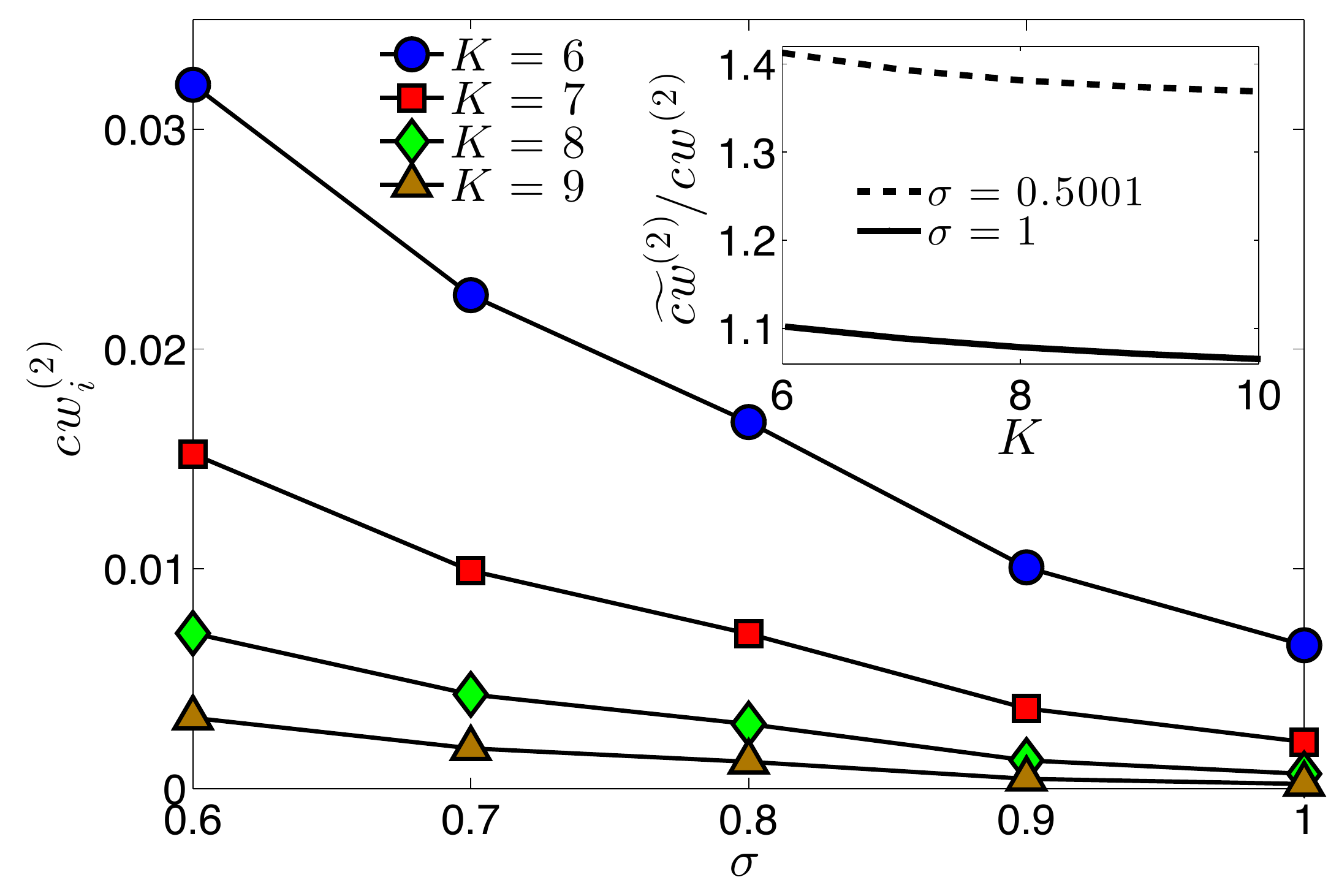}
	\caption{(Color online) Main plot: clustering $cw^{(2)}$ as a function of $\sigma$ and for different choices of $K$, as explained in the legend. The value of  $cw^{(2)}$ is monotonically decreasing with $K$ and with $\sigma$. Inset: ratio between the approximated value $\widetilde{cw}^{(2)}$ (calculated via Eq. $25$) and the exact values $cw^{(2)}$ (calculated numerically). More precisely, $\widetilde{cw}^{(2)}$ provides an upper bound for $cw^{(2)}$ and the approximation is better for large $\sigma$.}
	\label{fig:Onnela}
\end{figure}

The dependence on $\sigma$ of $cw^{(2)}$ and the goodness of the approximation provided by $\widetilde{cw}^{(2)}$ are visualised in Fig.~\ref{fig:Onnela}. In particular, $cw^{(2)}$ is relatively low and decreasing with $\sigma$. In fact, the definition (\ref{eq:c12}) takes into account the weights of all the links making up a triangle and the number of links between distant nodes (i.e., nodes loosely connected) is much larger than the number of links between close nodes (i.e., nodes tightly connected). Moreover, any weight is decreasing in $\sigma$ and, as a result, the overall clustering coefficient $cw^{(2)}$ is also decreasing in $\sigma$.

\subsection{The Hierarchical Ferromagnet  with noise: deterministic dilution}
We can allow for the presence of noise within the system
by assuming that links, whose weight is smaller than the noise level $T$, are ineffective (this mimics e.g., the fail or the unreliability of the link itself).
Therefore, despite the network we are considering is fully connected, when noise is present weaker weights, with $J_{ij} <T$, basically do not play any longer, as if they were missing \cite{ACG-PRE2011}. Since in the statistical mechanical analysis the noise level can be tuned arbitrarily \cite{NOI-JPA, NOI-NN}, it is crucial to understand how the overall network connection and clustering are accordingly modified.

\begin{figure}[h!]
\includegraphics[width=7.5cm]{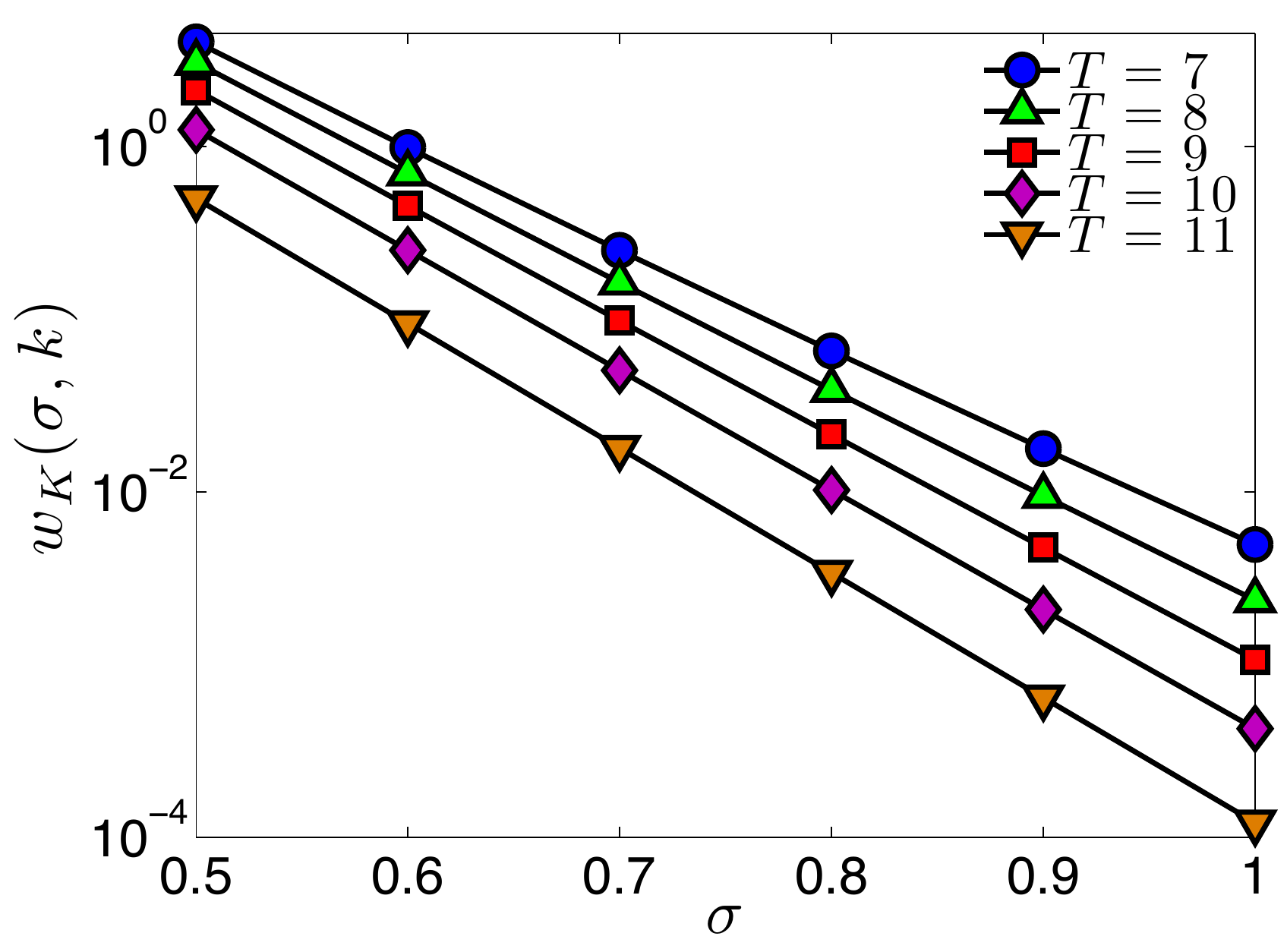}
\caption{(Color online) Weighted degree $w_K(\sigma,k)$ of nodes for the hierarchical graph where the presence of noise is mimicked by neglecting links displaying a weight smaller than $J(d)$, namely links connecting nodes at a distance larger than $d$. In this case $K=12$, $\sigma\in(0.5,1]$ and $T$ is taken varying in the interval $[7,11]$ (that means to neglet links such that $J(d)<T$). Notice that the higher values of $w$ are obtained in correspondence of low $\sigma$ and $d$; when $d=K=12$ all links are neglected.}
\label{fig:wvsTDyson}
\end{figure}

The analysis described in the previous section can be generalized in these terms.
For instance, the distribution $n_K(J)$ will exhibit a lower cutoff, being $n_K(J)=0$ for any $J<T$.
As for the weighted degree, $w_K(\sigma)$ (see Eq.~\ref{eq:W}) can be extended to $w_K(\sigma, k)$ reading as
\begin{eqnarray}
\label{eq:W2}
w_K(\sigma, k) &=& \sum_{d=1}^{k} 2^{d-1} J(d)=\\
\nonumber
& =& \frac{1}{4^{\sigma}-1} \left[ 4^{\sigma} \sum_{d=1}^k 2^{d(1-2\sigma)}- 4^{-k \sigma} \sum_{d=1}^k 2^d  \right]=\\
\nonumber
&=& \frac{4^{-\sigma k} \Gamma(2^{2\sigma-1},k)}{(2^{2\sigma-1}-1) (2^{2\sigma}-1)},
\end{eqnarray}
where $k=k(T)=1-\frac{1}{2\sigma}\log_2[T(4^{\sigma}-1)+4^{-K\sigma}]$, namely $k=\min_{i\in[1,K]}\{J(i)<T\}$, and $\Gamma(t,j)= 2^j-1+t-2^{j+1}t+2^jt^{j+1}, j\in[1,K]$.
Of course, by definition, $w_K(\sigma, K) \equiv w_K(\sigma)$.
These results are summarized in Fig.~\ref{fig:wvsTDyson}, where the behaviour of the weight of nodes is computed, as the level of noise $T$ and the parameter $\sigma$ are varied.

As for the clustering coefficient, we are interested in understanding whether, as the level of noise is increased, the giant component breaks into structure-less parts or it retains a large degree of clustering. The expression for the weighted clustering coefficients can be generalized into $cw^{(1,2)}(k)$ to account for the presence of some noise that impairs weak links. When $k(T)=K-1$, the $n(k) = N^2/4$ weakest links are neglected and the graph breaks down in two equal components of size $N/2$ which are a rescaled version of the original graph. Hence, for any node of each component  $cw^{(1)}(k=K-1)$ is still unitary. As noise is raised each component of the graph is further split and the resulting components all form weighted cliques.
Analogous arguments also hold for the degree of modularity.

On the other hand $cw^{(2)}(k)$, is quantitatively affected by the level of noise which further reduce its value. 

Therefore, even in the presence of noise, we can look at $\mathcal{G}$ as a clustered structure with a large degree of redundancy.

\subsection{The Hierarchical Ferromagnet as a Markov chain}

The graph modeling the hierarchical ferromagnet displays a countable set of nodes and finite weights, i.e. $J_{\textrm{min}} \leq J_{ij} \leq J_{\textrm{max}}$, for any couple $(i,j)$. Given such properties, upon proper normalization of weights $J_{ij} \rightarrow W_{ij} = J_{ij} / w_i$, the graph $\mathcal{G}(V,E)$ describes a Markov chain, where $V$ is the state space (each node $i$ represents a state) and $\textbf{W}$ is the transition matrix (see e.g., \cite{Norris-1994}).

We now focus on the pattern of couplings and check the stationary states, without any (direct) concern about spin dynamics: we will see, however, that the latter share several properties with those of this Markov chain \cite{NOI-JPA}.

Due to the symmetry among rows and columns (the summation over the rows equals $1$ as the summation over the columns) that the graph implicitly has, $\textbf{W}$ is not only stochastic, but even doubly stochastic. We also introduce a distribution \footnote{As such, $p$ must satisfy positivity, i.e. $\forall i\in V, 0 \leq p_i$, and normalization, i.e. $\sum_{i \in V} p_i=1$.} $p = (p_i: i \in V)$ on $V$ in such a way that the probability to find the random process in a state $i$ is given by $p_i$.
The evolution of the stochastic process is then provided by the following master equation
\be
p(t+1) = W p(t) \rightarrow \dot{p}(t) = W p(t) - p(t).
\ee
Therefore, the stationary distribution, referred to as $\pi$, satisfies $\pi = \textbf{W} \pi$, that is, $\pi$ coincides with the eigenvector $\phi_{\lambda_0}$ of $\textbf{W}$ corresponding to eigenvalue $\lambda_0=1$.
Due to the stochasticity of $\mathbf{W}$,  $\lambda_0=1$ is just the Perron-Frobenius eigenvalue of $\mathbf{W}$ and $\pi= \mathbf{e}/\sqrt{N}$, where all the $N$ entries of the vector $\mathbf{e}$ are equal to $1$.
      
The particular symmetry of $\textbf{W}$ allows to see that the states
\bea\nonumber
\phi_{\lambda_1} &=& (\underbrace{1, 1, \dots, 1}_{N/2}, \underbrace{ -1, -1, \dots, -1}_{N/2})/\sqrt{N},\\ \nonumber
\phi_{\lambda_2} &=& (\underbrace{1, \dots, 1}_{N/4}, \underbrace{ -1, \dots, -1}_{N/4},\underbrace{1, \dots, 1}_{N/4}, \underbrace{ -1, \dots, -1}_{N/4})/\sqrt{N},\\ \nonumber
\phi_{\lambda_3} &=& (\underbrace{1, \dots, 1}_{N/8}, \dots, \underbrace{ -1, \dots, -1}_{N/8})/\sqrt{N},
\eea
(and so on), are also eigenstates of $\textbf{W}$ and the related eigenvalues are
\bea
\nonumber
\lambda_1&=& \sum_{j=1}^{N/2} W_{1j} -\frac{N}{2} W_{1N}  = \frac{1}{w} \left( \sum_{j=1}^{N/2} J_{1j} -\frac{N}{2} J_{1N}  \right) \\
\label{eq:l1}
&=&\-\ 1 - \frac{N (2^{2\sigma} -2) (2^{2\sigma} -1)}{(2N)^{2 \sigma} - 2N (2^{\sigma} -1) +2^{2\sigma} -2} \approx 1 - \frac{2^{4 \sigma}}{N^{2\sigma -1} },\\
\nonumber
\lambda_2 &=& \sum_{j=1}^{N/4} W_{1j} -\sum_{j=N/4+1}^{N/2} W_{1j}  = \frac{1}{w} \left( \sum_{j=1}^{N/4} J_{1j} - \sum_{j=N/4+1}^{N/2} J_{1j} \right) \\
\label{eq:l2}
&=&  1 - \frac{N(2^{2 \sigma} -1)( 2^{2 \sigma-1} -1) (2^{2 \sigma} +2)}{(2N)^{2\sigma} -2N(2^{\sigma} -1) + 2^{2\sigma}-2} \approx 1 - \frac{2^{6 \sigma-1}}{N^{2 \sigma-1}}\\
\nonumber
\lambda_3 &=& \sum_{j=1}^{N/8} W_{1j} -\sum_{j=N/8+1}^{N/4} W_{1j}  = \frac{1}{w} \left( \sum_{j=1}^{N/8} J_{1j} - \sum_{j=N/8+1}^{N/4} J_{1j} \right) \\
\label{eq:l3}
&=&1 - \frac{N(2^{2\sigma}-1)(2^{6\sigma-2}-2)}{(2N)^{2\sigma} -2N(2^{2\sigma} -1) + 2^{2\sigma}-1} \approx 1 - \frac{2^{8 \sigma-2}}{N^{2 \sigma-1}}.
\eea
where the approximation in the last passages holds in the thermodynamic limit and we adopted the convention $1 = \lambda_0 \geq \lambda_1 \geq \lambda_2 \geq \dots \geq \lambda_N$. In general, one can see that $\lambda_l \approx 1 - 2^{2\sigma (l+1) - (l-1)}/N^{2 \sigma -1}$.


Incidentally, we notice that $\lambda_1$ is exactly the difference between the external fields acting on spins when their state is fixed as $S_i =1, \forall i$ and as $S_i = 1 , \forall i \leq N/2 \wedge S_i = -1 , \forall i > N/2$, respectively (as clearly the field acting on a node $i$ is $h_i = \sum_j J_{ij}S_j$). More generally, $\lambda_l$ corresponds to the field acting on spins in the $l$-th metastable state of the model \cite{NOI-PRL,NOI-JPA}.


Moreover, as one can see from Eq.~\ref{eq:l1},  $\lambda_1$ converges to $1$ in the thermodynamic limit and this suggests an ergodicity breaking for the stochastic process (which, in turn, mirrors ergodicity breaking in statistical mechanics too and hides the presence of several metastable states in the model thermodynamics \cite{NOI-PRL,NOI-JPA}). In fact, $\phi_{\lambda_0}$ and $\phi_{\lambda_1}$ generate a subspace such that any vector in this subspace (hence writable as a linear combination of $\phi_{\lambda_0}$ and $\phi_{\lambda_1}$) is an eigenvector of $\textbf{W}$ of the same eigenvalue $\lambda = 1$. In particular, 
\bea
\phi_{\lambda_0} + \phi_{\lambda_1} = (\underbrace{1, 1, \dots, 1}_{N/2}, \underbrace{ 0, 0, \dots, 0}_{N/2}) \sqrt{2/N},\\
\phi_{\lambda_0} - \phi_{\lambda_1} = (\underbrace{0, 0, \dots, 0}_{N/2}, \underbrace{ 1, 1, \dots, 1}_{N/2}) \sqrt{2/N},
\eea
corresponds to a stationary state localized on the left and on the right branches of the graph, respectively.
\newline
Actually, it is easy to see that even $\lambda_2$ and $\lambda_3$ converge to $1$ in the thermodynamic limit, although with slower rate. The degenerate eigenstate therefore allows for stationary distributions localized on smaller portions of the structure. Indeed, this can be proved to hold iteratively by including eigenstates of higher and higher order which allow eigenstates localized in smaller and smaller portions of the graph, provided that the hierarchical symmetry is fulfilled (thus this decomposition can not be pushed up indefinitely so to reach the lowest structures as dimers, because it would be unstable \cite{NOI-JPA,NOI-NN}).

This means that if we initialize the stochastic process in any node $i$, as long as $K$ is finite, the distribution describing the state of the graph will reach a stationary state broadened over the whole set of states $V$ with equal probability, that is, the dynamic process on $\mathcal{G}$ is ergodic. But, as $K \to \infty$ the system tends to be localized only on the subset of nodes $V_1=\{1,2,...,N/2\}$ if $i \leq N/2$ or on the subset $V_2 = V \setminus V_1$  if $i > N/2$. More precisely, if the system is initially prepared according to the distribution $\phi_{\lambda_0} + \phi_{\lambda_1}$, in the thermodynamic limit it will never reach any node $j>N/2$, hence ergodicity is broken (and, correspondingly in statistical mechanics, metastable states become stable \cite{NOI-PRL,NOI-JPA}). If we simply assume that $K \gg 1$, then any localized state belonging to, say, the portion $V_1$, displays a characteristic time scale before broadening over the whole structure. The larger the subspace and the longer the time-scale; the larger $K$ and the longer the time-scales. In the thermodynamic limit timescale diverges, conferring even to this (non-mean-field) ferromagnet a glassy flavour.

\subsection{The spectral gap of the Hierarchical Ferromagnet}

\begin{figure}[tbh]
	\includegraphics[width=8.5cm]{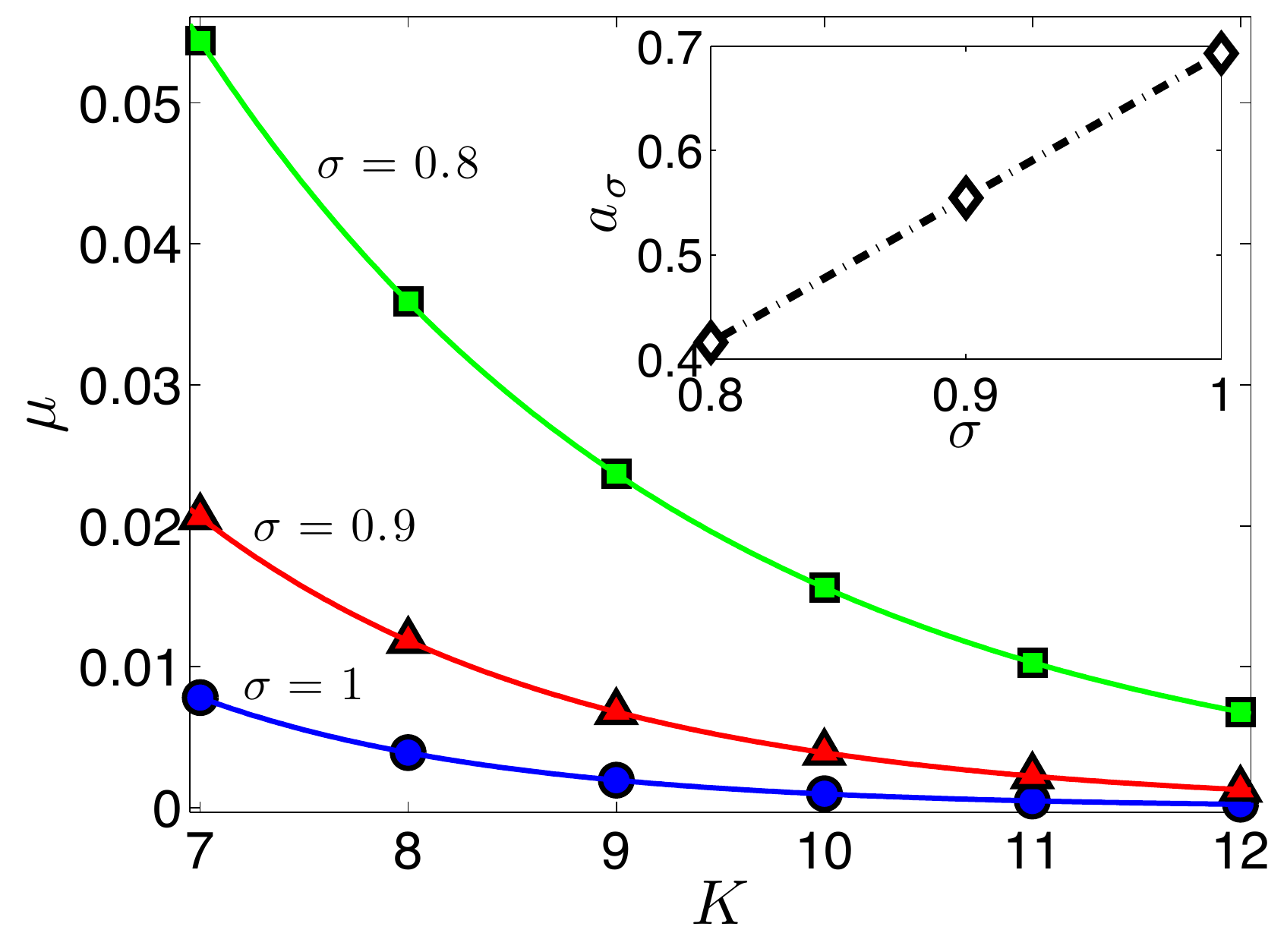}
	\caption{(Color online) Spectral gap $\mu$ as a function of $K$ (varying in $[7,12]$) and of $\sigma$ (with values $0.8$, $0.9$, $1$). As expected, the spectral gap decreases with the system size and with $\sigma$. Data from numerical calculations (symbols) are fitted (solid lines) via the function $y = \exp(- a_{\sigma}x)$. The dependence on $\sigma$ of the parameter $a_{\sigma}$, obtained from fitting procedures, is shown in the inset where the monotonic behaviour is highlighted.   
}
	\label{fig:SG}
\end{figure}

A close way to see the breakdown of ergodicity is to look at the spectral gap of the related Laplacian matrix. 
We already observed that the coupling matrix $\mathbf{J}$ is a block matrix, and every element $J_{ij}$ represents the weight of the link connecting the nodes $i$ and $j$. We can now introduce the Laplacian matrix, defined as $\mathbf{L}=\mathbf{Z}-\mathbf{J}$, where $\mathbf{Z}$ is a diagonal connectivity matrix such that $Z_{ii}=w$, $\forall i\in[1,N]$ . The eigenvalues $\mu_i$ of $\mathbf{L}$ satisfy
\beas
0 &=& \mu_0\leq\underbrace{\mu_1}_1 \leq\underbrace{\mu_2=\mu_3}_{2}\leq\underbrace{\mu_4=\mu_5=\mu_6=\mu_7}_{4}\leq\\ &\leq &\underbrace{\mu_8=...=\mu_{15}}_{8}\leq ...\leq\underbrace{...}_{N/2}\leq 2w,
\eeas
and we call spectral gap $\mu$ of $\mathbf{L}$ the smallest non trivial eigenvalue of $\textbf{L}$ \cite{munoz}. In particular, the smaller $\mu$ and the lower the number of links we need to cut so that the graph is divided in two independent blocks. In the hierarchical ferromagnet, we expect that this value tends to zero when the size of the system increases, obtaining the division of the network in two independent subgraphs, not interconnected. As depicted in Fig.~\ref{fig:SG}, $\mu$ goes to zero  exponentially with $K$ according to $f(K)=e^{-a_{\sigma}K}$. By fitting numerical data we find that the rate $a_{\sigma}$ decreases with $\sigma$ meaning that the higher the value of $\sigma$ and the lower the cost to fragment the graph.


We close with a remark. Once the size of the network fixed, the degree of modularity grows with $\mu$. Accordingly, we expect that the mean time for the Markov process (e.g. a random walker) to get broadened over the whole system grows with $\mu$. Therefore, consistently with \cite{Gallos-PNAS2007}, we find that modularity has a role in slowing down the transport process on a network.

\section{Graph generation in the hierarchical neural network}\label{sec:HHM}

Let us consider the hierarchical weighted graph $\mathcal{G}$ and let us generalize its coupling matrix $\mathbf{J}$ in order to account for the Hebbian prescription. This can be done following the so-called attribute approach: each node $i \in V$ is endowed with a set of attributes $\xi_i$ encoded by a vector of length $P$ whose entries are dichotomic and defined stochastically (see Eq.~\ref{eq:attribute}). The coupling $X_{ij}$, arising by comparing $\xi_i$ and $\xi_j$, is meant to mimic a learning process hence correlating/uncorrelating (i.e. strongly/poorly connecting).

The coupling matrix $\mathbf{X}$ is then used to modulate the former $\mathbf{J}$ in such a way that the final coupling matrix $\mathbf{Q}$ is given by the element-wise product
\be \label{eq:Q}
Q_{ij} = X_{ij} J_{ij},
\ee
for any couple $(i,j)$.
%
More precisely, recalling Eq.~\ref{HHM}, we have
\be \label{eq:Heb}
X_{ij} = \sum_{\mu=1}^P \xi_i^{\mu} \xi_j^{\mu},
\ee
which is also known as Hebbian rule in the neural-network context \cite{hebb}.
In this way, even close (according to the ultrametric distance) couples may  possible exhibit an overall null coupling if it occurs that the related entry in $\mathbf{X}$ is null.
Basically, $\mathbf{J}$ favors couples which are close according to the \emph{ultrametric} distance (defined on the set $\{ i \}$), while $\mathbf{X}$ favors couples which are close according to the \emph{Hamming} distance (defined on the set $\{ \xi \}$) \cite{Agliari-EPL2011}.

Notice that  $X_{ij}$ is a stochastic variable fulfilling a binomial distribution peaked at zero and with variance scaling linearly with $P$ \footnote{$X_{ij}$ can be looked at as the position reached by a one-dimensional simple random walk after $P$ steps.}. As both $X_{ij}$ and $J_{ij}$ are bounded we have
\ba
Q_{\textrm{max}} &=& J_{\textrm{max}}  X_{\textrm{max}}  = J(1) \times P = \frac{P(1 - 4^{-K \sigma})}{4^{\sigma} -1}, \\
Q_{\textrm{min}} &=& -Q_{\textrm{max}},\\
|Q|_{\textrm{min}} &=& J_{\textrm{min}}  |X|_{\textrm{min}} = 0.
\ea
where the third line derives from the fact that $Q$ is symmetrically distributed around $0$.

Moreover, as long as $P$ is large enough, we can write the following distribution for the coupling $Q_{ij}$
\begin{eqnarray} \label{eq:J}
\nonumber
&&P_{K,P}(Q_{ij}=q; \sigma) = P_{K,P}(X_{ij}= q/J_{ij} ; \sigma) \\
&&  =\frac{1}{\sqrt{2 \pi P} } \exp \left\{ - \frac{q^2 (4^{\s} -1)}{2P [4^{\s(1-d_{ij})} - 4^{-K \s}] } \right \},
\end{eqnarray}
where, exploiting the central limit theorem, we replaced the binomial distribution with a Gaussian distribution \footnote{Strictly speaking the convergence to a Gaussian distribution is better performing in the high storage regime only, namely where $P \sim N$; however, as we want to approach -trough this perspective- the topology of the hierarchical spin glass too, we allow ourselves in taking such an approximation.}.
\newline
The formalization just described can be properly extended to allow for correlation among string entries (e.g., see \cite{NOI-NN}) and for dilution in string entries (e.g. \cite{PRL}).

\begin{figure}[h!]
\includegraphics[width=9cm]{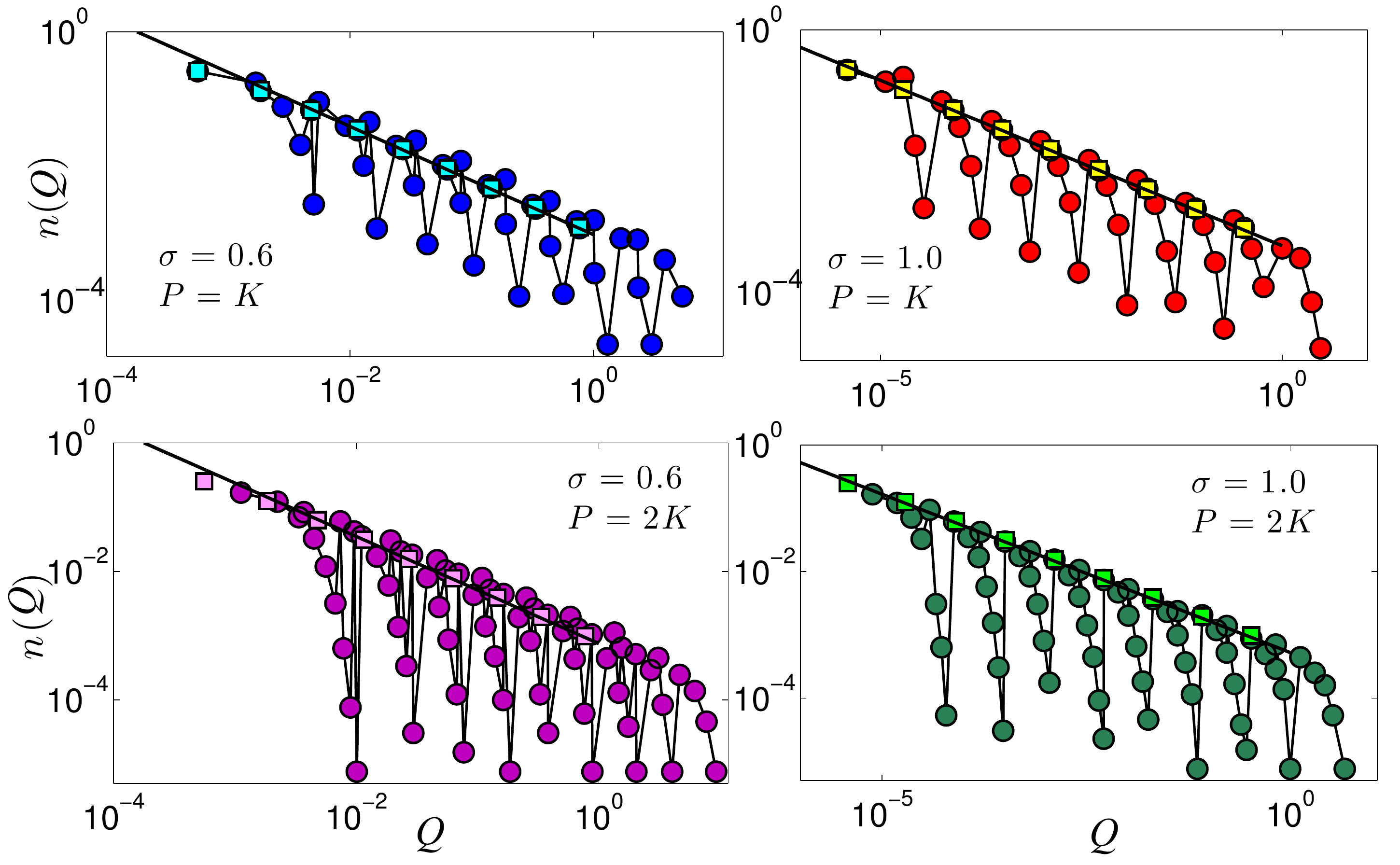}
 \caption{(Color online) Distribution $n_K(|Q|)$ for $K=9$ and different choice of the parameters $\sigma$ and $P$ as specified. Bullets represent data points for the graph generated by $\mathbf{Q}$ (see Eq.~\ref{eq:Q}),  squares represent the distribution one would obtain from the ultrametric contribution $\mathbf{J}$ only (see Eq.~\ref{eq:J}), and straight lines correspond to $y=J^{-1/(2\sigma)}/(2N)$ (see Eq.~\ref{eq:nJ}). }
\label{fig:distr2}
 \end{figure}

Here we focus on the simplest case (following Eqs.~\ref{eq:attribute} and ~\ref{eq:Heb}) and we start the investigation by looking at how the distribution of weights $n(Q)$ is affected by the modulation induced by $\mathbf{X}$. Results for several choices of the parameters $\sigma$ and $P$ are shown in Fig.~\ref{fig:distr2} (actually, due the symmetry of the distribution we can focus just on positive weights). With respect to the case analyzed in Sec.~\ref{sec:GD} and corresponding to the graph generated by the ultrametric contribution $\mathbf{J}$ only, here the set of possible values for weights $Q_{ij}$ is $2P+1$ or $2P$ times larger, according to the parity of $P$:
\beas
J_{ij} &\in& \{ J_1, J_2, \dots, J_K\} \rightarrow Q_{ij} \in \{0, \pm 2 J_1, \pm 4 J_1, \dots, \pm P J_1,\nonumber \\  &\pm& 2 J_2, \dots, \pm P J_K \} \;\; \textrm{if} \;\; P  \;\; \textrm{even}\\
J_{ij} &\in& \{ J_1, J_2, \dots, J_K\} \rightarrow Q_{ij} \in \{\pm J_1, \pm 3 J_1, \dots, \pm P J_1,\nonumber\\  &\pm& J_2, \dots, \pm P J_K \} \;\; \textrm{if} \;\; P \;\; \textrm{odd}.
\eeas
As a result, focusing on $P$ odd to fix ideas, $n(Q)$ is enveloped by the power law $Q^{-1/(2\sigma)}$, which matches the values $J_1, J_2,...,J_K$, and such values are also accompanied by other $P-1$ values whose occurrence follows a binomial distribution.

Notice that a large $P$ implies a broader distribution; similarly, a small $\sigma$ implies a larger support.
Therefore, we expect that the pattern of $\mathbf{Q}$ is still reminiscent of the hierarchical underlying structure, yet it is perturbed and the extent of such perturbation is more evident when $P$ is large (Fig.~\ref{fig:carpet}).

 \begin{figure*}
\includegraphics[width=1.0\textwidth]{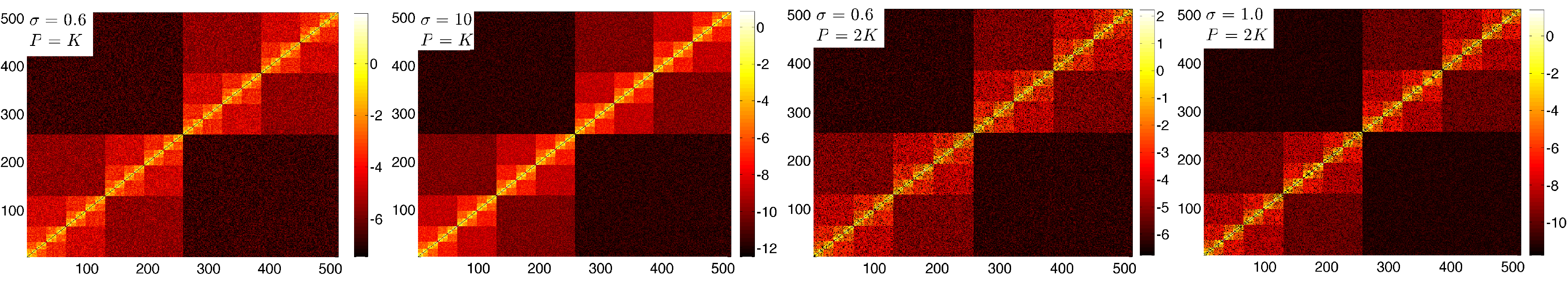}
 \caption{(Color online) Patterns of weights $Q_{ij}$ for $K=9$ and different choices of $\sigma$ and $P$ as reported; each pattern refers to a different colorbar. Notice that for a better visibility we plotted $\log(|Q_{ij}|)$, and that, for a given $\sigma$, by increasing $P$ the pattern gets more noisy; moreover, small values of $\sigma$ imply a larger support. This figure mirrors Fig.~\ref{fig:distr2}.}
 \label{fig:carpet}
 \end{figure*}

\begin{figure}
\includegraphics[width=8.2cm]{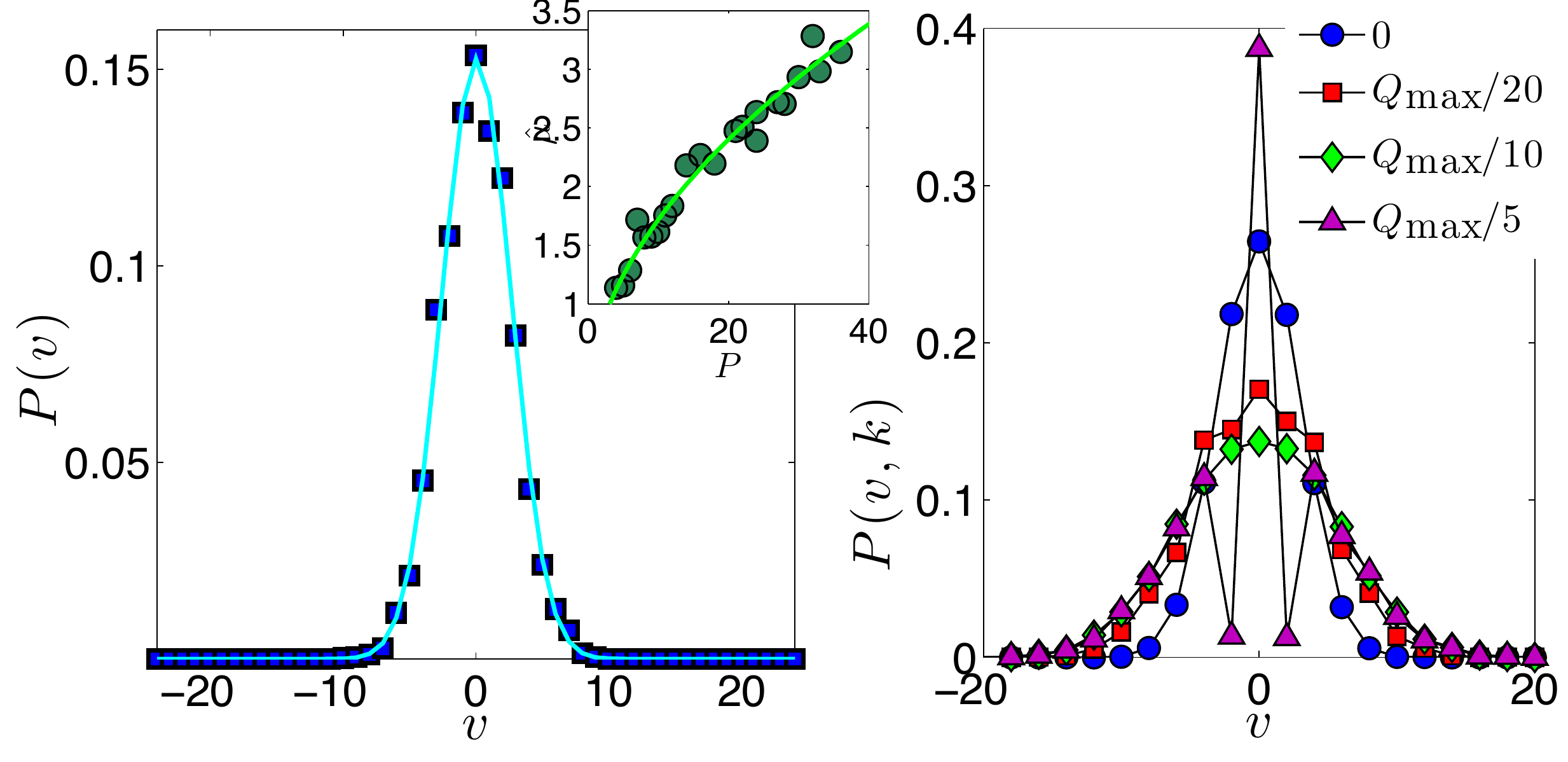}
 \caption{(Color online) Left panel: in the main plot we show the distribution of weighted degrees $P(v)$ for a system with $K=12$, $P=2K$ and $\sigma=1$. The distributions pertaining to different choices of parameters $K$ and $P$  were fitted with a central normal distribution with standard deviation $\hat{\mu}$. The best-fit values for $\sigma$ are shown versus $P$ in the inset where the solid line represents the function $y \sim \sqrt{x}$. Right panel: distribution $P(v,k)$ of generalized degree $q$ in a network with $K=12$, $P=2K$, $\sigma=1$ and different level of noise $k$. In particular, the set of noise is such that links corresponding to weights $Q_{ij}$ smaller than a given threshold are cut; the threshold considered are represented with different symbols as explained by the legend.}
 \label{fig:W_Hop}
 \end{figure}

We now calculate the weighted degree of node $i$ defined as
\be
w_i = \sum_{j=1}^N Q_{ij} = \sum_{j=1}^N J_{ij} X_{ij}.
\ee
Differently from the hierarchical ferromagnet model, here the strict homogeneity among nodes is lost and, in general, $w_i$ is site dependent.
We can therefore estimate the distribution $n(w)$ of weighted degrees: recalling  Eq.~\ref{eq:W} and that $X_{ij}$ is normally (at least as long as $P$ is sufficiently large) distributed with variance $P$, we expect that $v=w_i/w$ (with $w=\sum\limits_{i=1}^{N}J_{ij}$) is normally distributed with variance scaling with $P$. This is indeed checked numerically as shown in Fig.~\ref{fig:W_Hop}.

In this case, since $w_i$ is a random variable, we are interested in its mean and variance with respect to the random variables $\xi$. Recalling that
\be
w_i=\sum_{i\neq j} J_{ij}X_{ij}, \text{ with } J_{ij}X_{ij}=J(d_{ij})\sum\limits_{\mu=1}^{P}\xi_i^{\mu}\xi_j^{\mu},
\ee
one can see that the expected (according to the distribution in Eq.~\ref{eq:attribute}) value of $w_i$ is $\mathbb{E}(w_i)=0$ and, computing the variance of $w_i$, we obtain
\beas
& &\Var(w_i)=\sum\limits_{i\neq j}J(d_{ij})\Var\left[\sum\limits_{\mu=1}^{P}\xi_i^{\mu}\xi_j^{\mu}\right]=\sum_{i\neq j} P \times J(d_{ij})=\\\nonumber
&=&  \frac{P}{2(4^{\sigma} -1)} \left[ \frac{(2N)^{2 \sigma} -N(3 \times 2^{2 \sigma-1} -1) +2^{2\sigma} -1  }{N^{2\sigma}(2^{2 \sigma-1}-1)}  \right],
\eeas
where $P$ is the variance of $X_{ij}$.

\begin{figure*}
\includegraphics[width=15cm]{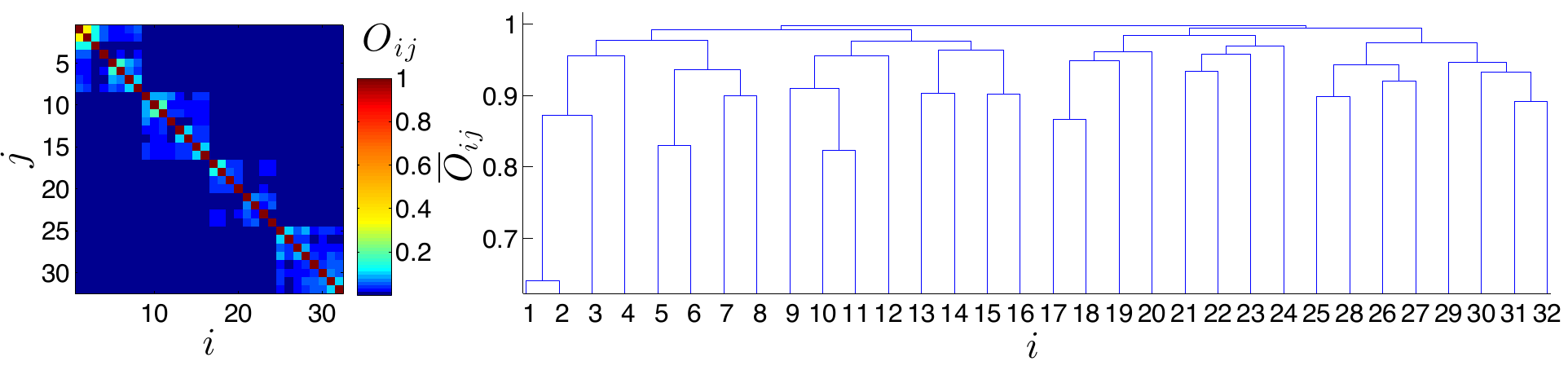}
	\caption{(Color online) Representation of modularity between nodes using the topological overlap matrix $\mathbf{O}$ for fixed $K=5$, and $\sigma=0.9$. 
	Left panel: Matricial representation of overlaps obtained via Eq.~\ref{eq:GTOM2}. Different colours represent different values of overlap, as explained by the colorbar on the right. Due to the presence of random variables in the construction the coupling matrix $\mathbf{Q}$, we  can see partial loss of regularity in the structure of the matrix. 
	Right panel: dendrogram showing the dissimilarity between nodes in the graph: nodes at distance $d_{ij}=1$ (e.g., node $i=3$ and $j=4$) typically display high overlap (hence have low dissimilarity) with respect to those at distance $d_{ij}=2$ (e.g., $i=13$ and $j=15$), up to the maximum distance $d_{ij}=5$ (e.g., $i= 1$ and $j=24$), underlying the ultrametric structure of the network. However, differently from the case of hierarchical ferromagnet, this case is irregular and overlaps, especially between nodes at close distance, are broadly distributed. }
	\label{fig:GTOMHop}
\end{figure*}

The next step is to evaluate the degree of modularity. Differently from the hierarchical ferromagnet, where there is a perfect homogeneity in the weight of nodes, here we expect the regular ultrametric structure to be perturbed by the stochastic factor $\mathbf{X}$. The generalised overlap matrix $\mathbf{O}$ (see Eq.~$\ref{eq:GTOM2}$ for its extended formula) is computed and shown in Fig.~\ref{fig:GTOMHop}, along with a dendrogram plot capturing the dissimilarity (see Eq.~\ref{eq:disGTOM}) between nodes. For the realization considered the highest values of overlap are still obtained for dimers, yet the resulting structure is not fully regular as previously found for the hierarchical ferromagnet and two nodes at distance $1$ may in principle exhibit a relatively large dissimilarity.

We conclude this section stressing that, in the Hopfield network, the presence of $P$ random vectors $\xi_i^{\mu}$, $\mu=1,...,P$ peaked at zero  implies that 
it is no longer possible to establish that $Q(d_{ij})>Q(d_{hk})$ when $d_{ij}>d_{hk}$ and this is the cause of the loss of a regular structure in the overlap measure shown in Fig.~\ref{fig:GTOMHop}.

\section{Graph generation in the hierarchical spin glass}

\begin{figure}[h!]
\includegraphics[width=7.5cm]{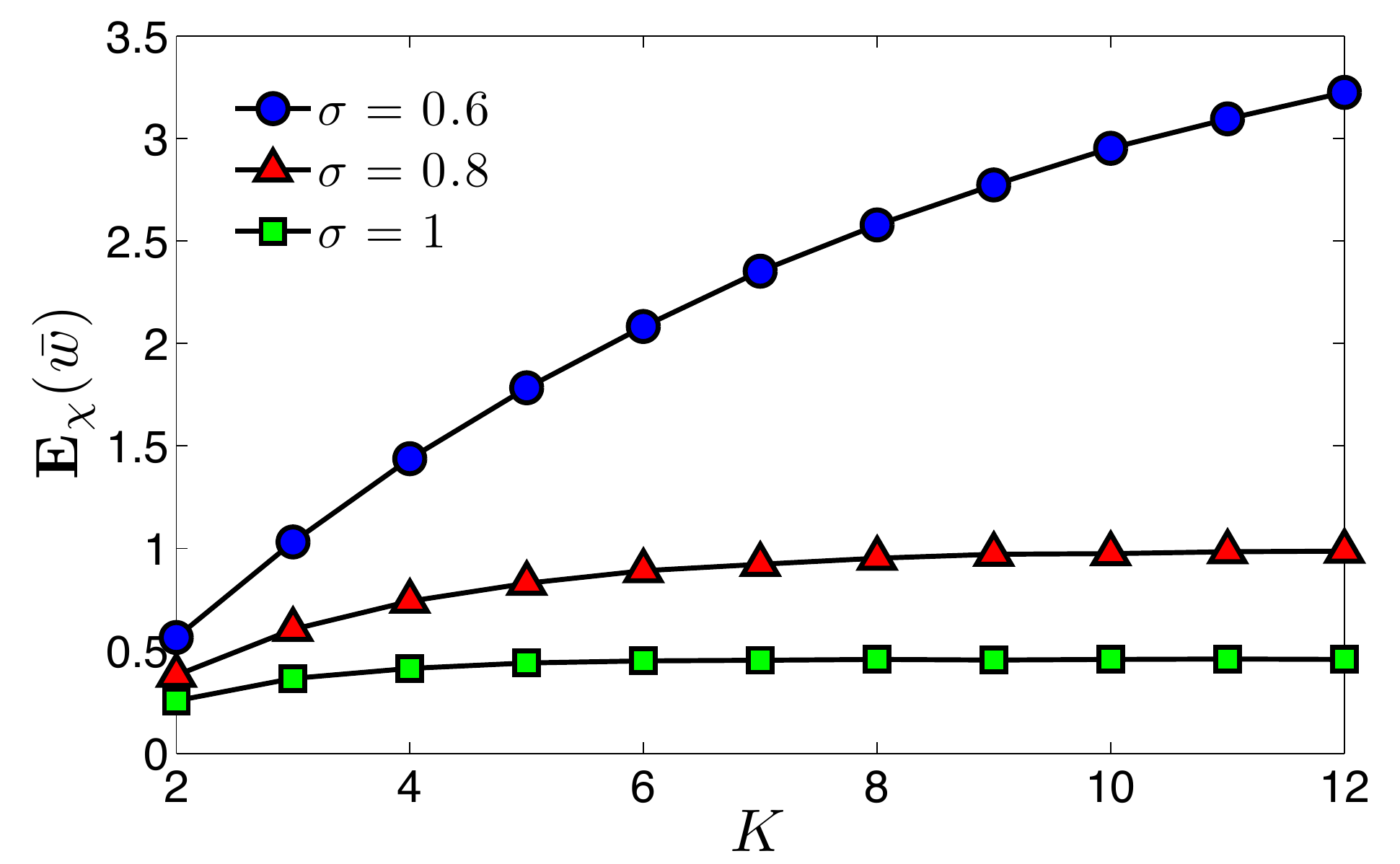}
 \caption{(Color online) Plot of $\mathbb{E}_{\chi_{ij}}[\overline{w}]=\mathbb{E}_{\chi_{ij}}[\frac{\sum_{i=1}^{N}w_i}{N}]$ as a function of the system size ($N=2^K$) and of $\sigma$ (as explained in the legend). To realize it, 200 realization of $\chi_{ij}$ were produced, and, for each, we obtained $w_i, \forall i\in[1,N]$. Then, the algebraic mean value over the realizations was computed, with fixed $K$. As expected, the largest values  of $w$ are obtained for the highest values of $K$ and for the lowest values of $\sigma$. }
 \label{fig:wvsKSK}
 \end{figure}

This section is devoted to the study of the generation of the weighted graph $\mathcal{G}$ in the case of hierarchical spin glass. As introduced in Sec.~II, in this case the couplings among spins are defined as quenched variables drawn from a standard centered Gaussian distribution $\mathcal{N}[0,1]$. This means that we can write
\be \label{eq:QQ}
Q_{ij}=\chi_{ij}J_{ij}=\chi_{ij}\frac{4^{\sigma(1-d_{ij})}-4^{-K\sigma}}{4^{\sigma}-1},
\ee
where $\chi_{ij}$ are independent, centered Gaussian variables.
Due to the contribution of $\chi_{ij}$ in the definition (\ref{eq:QQ}), the weight of nodes is site dependent. More precisely, we have that the expected value of $w_i$ is $\mathbb{E}_{\chi_{ij}}[w_i]= 0$ and its variance reads as
\beas
&&\Var_{\chi_{ij}}[w_i]=\Var_{\chi_{ij}} \left [\sum_{i\neq j}\chi_{ij}J_{ij} \right]=\\
&=& \frac{(N-1)}{2(4^{\sigma} -1)} \left[ \frac{(2N)^{2 \sigma} -N(3 \times 2^{2 \sigma-1} -1) +2^{2\sigma} -1  }{N^{2\sigma}(2^{2 \sigma-1}-1)}  \right],
\eeas
where we used that a linear combination of random Gaussian independent variables is still a Gaussian variable, with variance equal to the sum of variances of the variables. Numerical results for the average value $\mathbb{E}_{\chi_{ij}}[\overline{w}]=\mathbb{E}_{\chi_{ij}}[\frac{\sum_{i=1}^{N}w_i}{N}]$ are shown in Fig.~\ref{fig:wvsKSK}.


\begin{figure*}
\includegraphics[width=15cm]{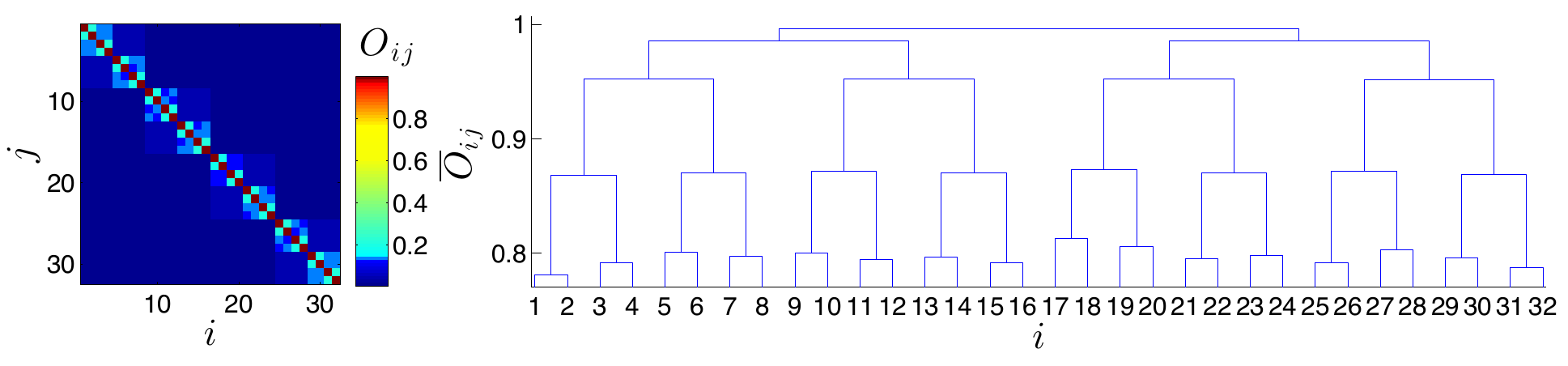}
\caption{(Color online) Representation of modularity between nodes using topological overlapping measure when $K=5$, $\sigma=0.9$. 
 Left panel: Matricial representation of overlaps using obtained via Eq.~\ref{eq:GTOM2}. Different colors represent different values of overlap, as explained by the colourbar on the right. Due to the presence of Gaussian random variables in the construction the coupling matrix, we can see again its random behavior,  but the network anyway results more regular than the hierarchical neural network.
Right panel: dendrogram showing the dissimilarity between nodes in the graphs: nodes at distance $d_{ij}=1$ (see for instance node $i=1$ and $j=2$, at the lower level of the dendrogram) have higher values of overlap (hence have lower values of dissimilarity) with respect to those at distance $d_{ij}=2$ (second level from the bottom of the dendrogram), up to the maximum distance $d_{ij}=5$ (first level on the top). As shown for the hierarchical neural network, also the hierarchical spin-glass partially loose the symmetric structure for the presence of random variables that contribute to tune the elements of adjacency matrix. }
\label{fig:SK_5_09}
\end{figure*}

We also checked the modularity of this networks by exploiting again the generalised topological overlap matrix \cite{GTOM} given by Eq.~\ref{eq:GTOM2}. In this case we obtain a more regular structure with respect to the hierarchical neural network, due to the presence of random quenched Gaussian variables.
This  perfectly matches with the definition of the model: the presence of the random variables $\chi_{ij}$, that contribute to construct the coupling matrix, introduces a random component that affects the overlap between dimers, squares, octagons ect., as depicted in Fig~\ref{fig:SK_5_09}.\\

The remarks highlighted for the neural network model (see the conclusion of the previous section) remain valid for the hierarchical spin-glass model as well: the presence  of weights on links depending on random variables leads to a loss of symmetry in the structure of the network: the links favoured by the ultrametric distance are not necessarily the same favoured by the random coupling $\chi_{ij}$.\\

%
%
%
%
%
%

\section{Conclusions and outlooks}\label{sec:conclusions-and-outlooks}

In the last decade hierarchical networks have been found to play a crucial and widespread role in natural phenomena \cite{barabba,moreno}, particularly in biological systems \cite{neurone,protein}.
Furthermore, these structures turn out to be also quasi-tractable in statistical mechanics \cite{dyson}, even when glassiness is present \cite{NOI-PRL,NOI-JPA,castellana-jsp,castellana-jsp2,castellana-prl,castellana-pre,dyson}, hence triggering further studies of their properties.
\newline
In this work we discuss the topological features of three hierarchical models, each describing a different rule for generating  couplings among nodes: Herarchical Ferromagnet (HFM), Hierarchical Neural Network (HNN) and Hierarchical Spin Glass (HSG).
In particular, we show that the subtle metastabilities exhibited by HFM (see e.g. \cite{NOI-JPA,NOI-PRL}) can also be evidenced in terms of ergodictiy breakdown for Markov processes defined on the hierarchical weighted graph embedding the spin system.

More precisely, the graph could be considered as a Markov chain, where the state space is the set of nodes, and entries in the transition matrix are constituted by the distances between nodes, upon a proper normalization: the breakdown of ergodicity is thus depicted by the divergence of the mixing time, mirroring the results obtained via the statistical mechanical route. 

Further, these structures also exhibit high clustering (at least according to the definition (\ref{eq:c12})) and modularity which are two important properties well-evidenced in many real systems \cite{Barrat-PNAS2004,Newman-2010}.

Analogous analyses were carried out for HNN and HSG. As expected (because now quenched disorder is introduced in the system), differences were shown to exist between the two of them and the HFM: the most important is the loss of the symmetric hierarchical structure of weights of the links, due to the presence of random variables that contribute to create the coupling matrix. For the HNN, the Hebbian rule leads to a binomial distribution for the weighted degree of nodes, that is peaked at zero and with variance scaling with $P$. For the HSG, the Gaussian term leads to a weighted degree peaked at zero and with variance scaling with $N$.
In general, the Hebbian contribute induces a smaller broadness but a larger noise. 

These investigations deepen the strong connection between the thermodynamic behaviour of a statistical mechanics model and the topological properties of the underlying structure. The ultimate goal is to contribute to the description of non-mean-field systems. 

\appendix
\section{Notes about modularity}\label{sec:appendix}

In this section we deepen the analysis of the network modularity introduced in Sec.~\ref{sec:GD} for the hierarchical ferromagnet.

\begin{figure}
\includegraphics[width=8cm]{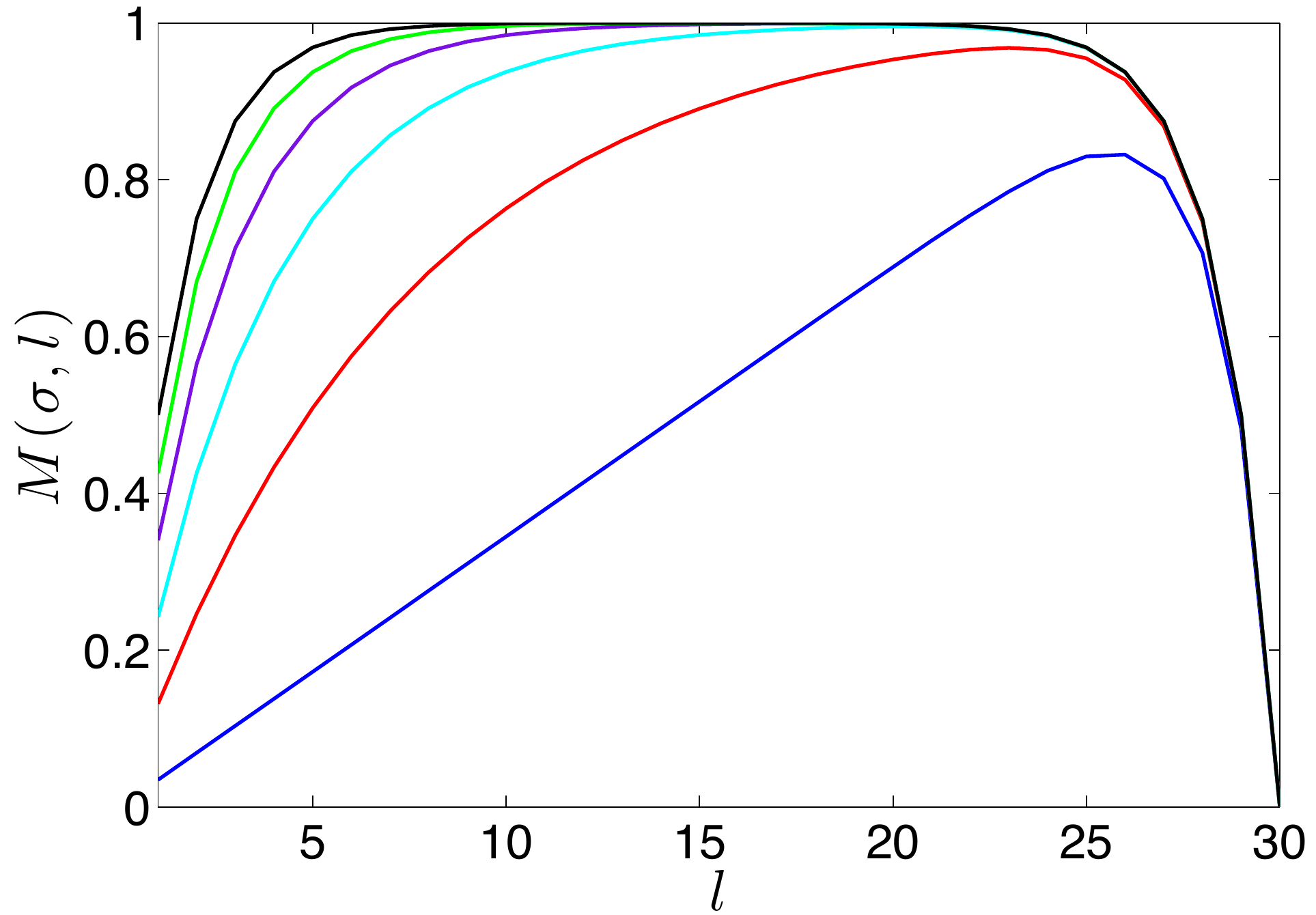}
\caption{(Color online) Degree of modularity $M(\sigma,l)$ for $K=30$ and different values of $\sigma$. Curves in different colors correspond to different values of $\sigma$: as $\sigma$ is varied from $0.5$ to $1.0$ the peak moves from right to left; see Eq.~\ref{eq:mod2}.}
\label{fig:mod2}
\end{figure}

In order to figure out which is the most effective partition, we can apply the formula introduced in \cite{Newman-2010} (suitable for weighted graphs as well) which measures the degree of modularity $M$ for a given modular subdivision, chosen a priori, where each node $i$ is associated to a module $m_i$ out of $\bar{m}$, i.e. $c_i =1,...,\bar{m}$. More precisely,
\be \label{eq:mod}
M=\frac{1}{m} \sum_{ij} \left[ J_{ij} - \frac{w_i w_j}{m} \right] \delta(c_i,c_j),
\ee
where $m=\sum_{i}w_i$.
In particular, exploiting the homogeneity of the hierarchical graph we can write Eq.~(\ref{eq:mod}) in a simpler form as
\be \label{eq:mod2}
M(\sigma,l)=\frac{1}{Nw} \sum_{i<j} \left[ J_{ij} - \frac{w}{N} \right] \delta(c_i,c_j).
\ee

According to the different modular subdivision, we can calculate the resulting $M$, and, in general, with communities made of $2^l$ nodes we have
\begin{eqnarray} \label{eq:mod2}
M(\sigma, l)&=&\frac{1}{Nw}   \sum_{d=1}^l \left[ J(d) - \frac{w}{N} \right]2^{l+d-1} \\
\nonumber
&=& \frac{2^l}{2^K}\Big(\frac{w_l(\sigma)}{w_K(\sigma)}+\frac{2^l-1}{2^K}\Big)=\\
&=& \frac{t^K}{t^l}\frac{\Gamma(t,l)}{\Gamma(t,K)}+\frac{2^l(2^l-1)}{2^{2K}},\nonumber
\end{eqnarray}
where we posed $t=2^{2\sigma-1}$ and where
\begin{eqnarray}\nonumber
w_l(\sigma)&=&\sum_{d=1}^{l}2^{d-1}J(d);\\ \nonumber
\Gamma(t,j)&=& 2^j-1+t-2^{j+1}t+2^jt^{j+1}, j\in[1,K].\label{F}
\end{eqnarray}
As shown in Fig.~\ref{fig:mod2}, the function $M(\sigma,l)$ exhibits a peak at a value $l$ approaching $k/2$ as $\sigma \to 1$. This means that the most effective modular partition (according to Eq.~\ref{eq:mod}) is the one where the graph is divided in a relatively small number of clusters, but for large $\sigma$ (namely where the hierarchy is less important, see Eq. \ref{eq:ABrho}) this number gets smaller.

Finally, we introduce an alternative formulation for extending the formula introduced in \cite{GTOM} and reported in Eq.~\ref{eq:GTOM}. In fact, exploiting the discreteness of the entries of the coupling matrix $\mathbf{J}$ we can write 
\be \label{eq:GTOMprime}
O^{\prime}_{ij} = \frac{\sum_{l=1}^K | N_l(i) \cap N_l(j) | J(l)}{ \min \{ w_i, w_j\}},
\ee
where $N_l(i)$ is the number of links with coupling $J(l)$ stemming from node $i$. 
In particular, the expression in Eq.~\ref{eq:GTOMprime} can be applied to the hierarchical ferromagnet, obtaining the following 
\begin{equation}
O^{\prime}_{ij}=\frac{(2N)^{2\sigma}2^{d(1-2\sigma)}+2N(1-2^{2\sigma})+2^{d}(2^{2\sigma}-1)}{(2N)^{2\sigma}-2N(2^{2\sigma}-1)+2^{2\sigma}-2},
\end{equation}
where to ligthen the notation we posed $d=d_{ij}$.
In the thermodynamic limit $O^{\prime}_{ij} \sim 2^{-d(2\sigma-1)}$, namely the similarity between two nodes decreases exponentially with their distance.

\section*{Acknowledgments}
GNFM-INdAM, INFN and Sapienza Universit\`a di Roma are acknowledged for partial financial support.


\end{document}